\documentclass[%
 aip,cha,
 amsmath,amssymb,
preprint,%
floatfix,
author-numerical,%
]{revtex4-1}

\usepackage{graphicx}
\usepackage{dcolumn}
\usepackage{bm}

\begin{document}

\preprint{AIP/123-QED}

\title{Community detection analysis in wind speed-monitoring systems using mutual information-based complex network}

\author{Mohamed Laib}
 \email{Mohamed.Laib@unil.ch}
  \author{Fabian Guignard}
 \author{Mikhail Kanevski}
 \affiliation{IDYST, Faculty of Geosciences and Environment, University of Lausanne 1015, Switzerland.} 
\author{Luciano Telesca} %
 \affiliation{CNR, Istituto di Metodologie per l'Analisi Ambientale, 85050 Tito (PZ), Italy.}


\date{\today}

\begin{abstract}
A mutual information-based weighted network representation of a wide wind speed monitoring system in Switzerland was analysed in order to detect communities. Two communities have been revealed, corresponding to two clusters of sensors situated respectively on the Alps and on the Jura-Plateau that define the two major climatic zones of Switzerland. The silhouette measure is used to evaluate the obtained communities and confirm the membership of each sensor to its cluster.

\end{abstract}

\keywords{Wind, Weighted network, Mutual information, Community detection, Time series}
\maketitle

\begin{quotation}
Since the installation of dense meteorological monitoring systems made available a huge amount of data, investigating the properties of meteo-climatic parameters has become challenging to understand the mechanisms underlying climatic systems. Complex networks represent an important theoretical framework that helps to describe and understand the interaction among meteo-climatic parameters concomitantly measured by sensors of a very dense monitoring system.  This work proposes a mutual information-based network to study the interaction between the wind speed series measured by a meteorological monitoring system in Switzerland, characterized by so diverse topographies. Applying a multilevel community detection method, two clusters of wind stations were identified, matching the two main climatic zones of Switzerland. The results of this study suggest new methodological approaches to investigate wind speed time series.

\end{quotation}

\section{Introduction}
\label{intro}
Over the past years, more and more data have been being collected at ever higher frequency that developing efficient pattern-detection methods and data-mining techniques has become very crucial to identify a few highly informative features. In this context, one of the most relevant examples is given by high-dimensional (multiple) time series that originate from constituent units of large systems characterized by inner interactions. 

The cooperative behaviour within a complex system involving relationships among its constituent units can be effectively described by networks, where the interactions among the constituents (or nodes of the network) are represented by links. The topology of the network, which coincides with the topology of such interconnections or links, is in itself complex \cite{Newman2003, Boccaletti2006}. These networks show a certain organization at a mesoscopic level, which is intermediate between the microscopic level that involves the single constituent units and the macroscopic level that involves the entire system as a whole. This mesoscopic level reflects the modular organization of the system, characterized by the existence of interconnected groups where some units are heavily linked with each other while, at the same time, are less correlated with the rest of the network. These interconnected groups are generally featured as communities \cite{Girvan2002, Fortunato2010}. Detecting such communities represents an important step in the dynamical characterization of a network, because it could reveal special relationships between the nodes that may not be easily detectable by direct empirical tests \cite{Lancichinetti2008}, this helps to a better understanding of the characteristics of dynamic processes that take place in a network.

The use of complex networks to understand the interactions characterizing a climatic system has been growing in the last years \cite{Tsonis2006, Tsonis2008, Donges2009a, Donges2009b, Gozolchiani2008}, and various approaches have been used in  constructing the related networks \cite{Tsonis2004, Yamasaki2008, Tsonis2008, Donges2009b, Steinhaeuser2009}. Further, complex network offer a new mathmatical modelling approach for non-linear dynamics \cite{Donner2019} and for climatological data analysis \cite{Donner2}.

Among the meteo-climatic parameters, wind is an important factor that influences the evolution of a climatic system; several studies have  been devoted to understand better its time dynamics by using several methods, like extreme value theory and copula \cite{DAMICO2015}, machine learning algorithms\cite{Treiber2016}, visibility graph analysis\cite{PIERINI2012}, Markov chains models \cite{KANTZ2004}, fractal \cite{DEOLIVEIRASANTOS2012, Fortuna2014}, multifractal analysis \cite{Telesca2016, Garcia2013}. 

The topological properties of wind systems have been a focus of investigation only in the very recent years. Laib et al. \cite{Laib2018a} studied the long-range fluctuations in the connectivity density time series of a correlation-based network of high-dimensional wind speed time series recorded by a monitoring system in Switzerland. They found that the daily time series of a connectivity density of the wind speed network is characterized by a clear annual periodicity that modulates the connectivity density more intensively for low than high absolute values  of the correlation threshold. 
Laib et al. \cite{Laib2018b} analysed the multifractality of connectivity density time series of the wind network and found that the larger multifractality at higher absolute values of thresholds could be probably induced by the higher spatial sparseness of the linked nodes at these thresholds.

Considering the topographic conditions of Switzerland and its wide-spread wind monitoring system, it is challenging to investigate the topology of the wind network in terms of existence of network communities, and to check if these communities match with the topography of the territory. 

To this aim, the edges of the network (the links between any two stations of the wind system, which are the nodes of the network) are weighted by the mutual information between the wind time series recorded at each station. The mutual information, which quantifies the degree of non-linear correlation between two time series, has been already used to construct seismic networks\cite{Jimenez2013}, global foreign exchange markets\cite{Cao2017}, prediction of stock market movements\cite{Kim2017}.

\section{Data and network construction}
\label{sec:1}
\begin{figure}
\centering
\includegraphics[width=\linewidth]{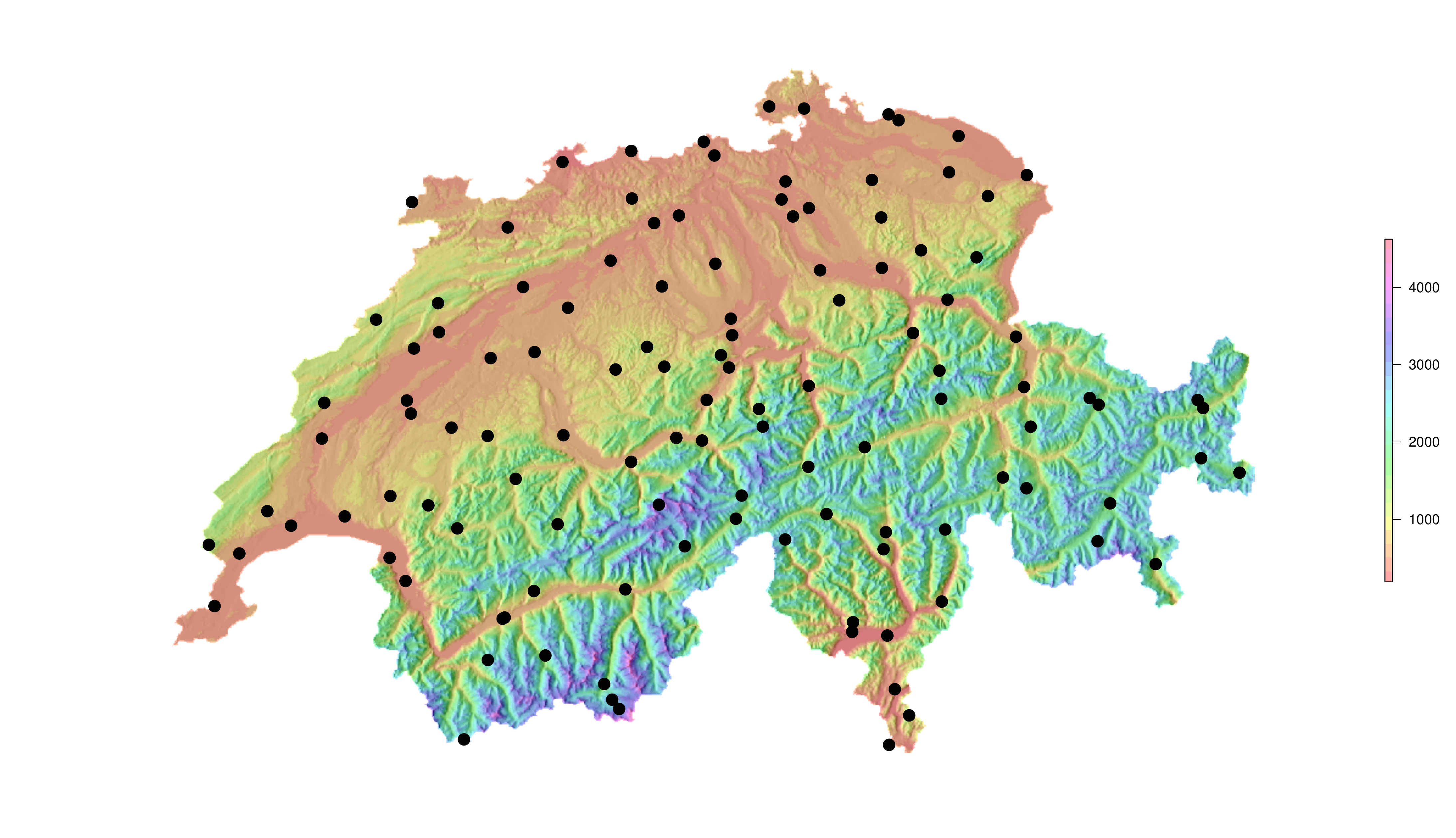}
\caption{Study area and location of measuring stations.}
\label{fig1}  
\end{figure}

\begin{figure}
\centering
\includegraphics[width=\linewidth]{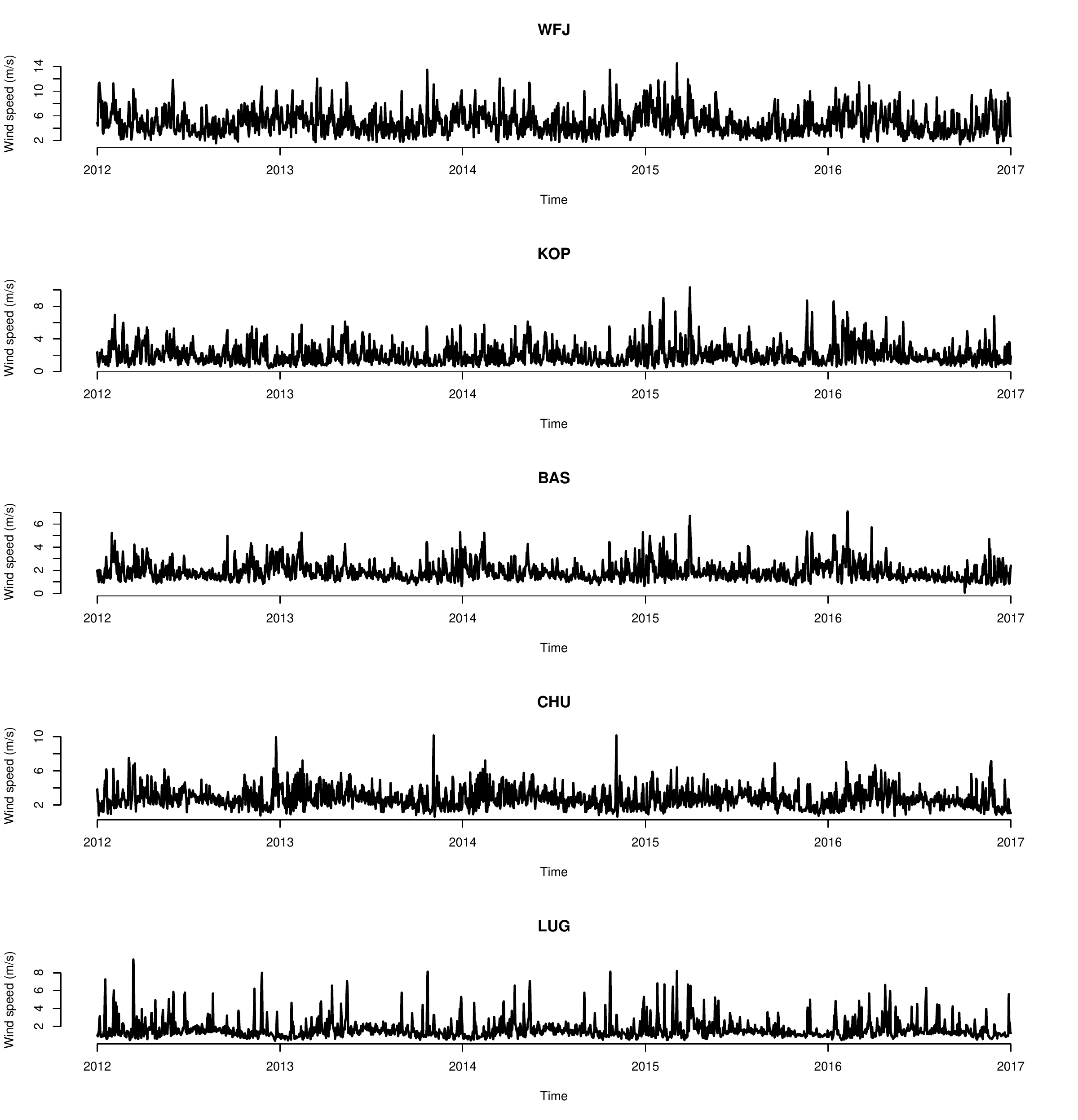}
\caption{Some example of daily wind time series.}
\label{fig2}  
\end{figure}

The data used in this work consists of daily mean wind speed, collected from 119 measuring stations from 2012 to 2016 by SwissMetNet, which is one of the weather monitoring systems in Switzerland covering almost homogeneously all the Swiss territory (Fig. \ref{fig1}). Fig. \ref{fig2} shows, as an example, some of the measured wind speed series. 

To  construct the network, the mutual information was used as a metric to weight the edges between the nodes:

\begin{equation}
\label{MIeq}
I(X,Y)=\sum_{x \in X} \sum_{y \in Y} p(x,y)log\left( \frac{p(x,y)}{p(x)p(y)}\right) 
\end{equation}
where $X$ and $Y$ are two different random variables (wind time series), $p(x)$ and $p(y)$ are respectively their probabilities, while $p(x,y)$ is their joint probability.

Mutual information is a measure of the amount of information that one random variable contains about another random variable \cite{MIbook}.

It can be shown that Eq. \ref{MIeq} can be written as follows\cite{MIbook}
\begin{equation}
I(X,Y) = D(p(x,y) \parallel p(x)p(y))
\end{equation}
where $D$ is the Kullback-Leibler divergence, which is a dissimilarity measure between two probability distributions. 

Thus, the mutual information can be seen as the departure of the joint probability $p(x,y)$ from the product of the two marginal probabilities $p(x)$ and $p(y)$. We can easily show that $I(X,Y) \geqslant 0$ with equality if and only if $X$ and $Y$ are independent \cite{MIbook}. Consequently, the higher the mutual information, the stronger the dependence between $X$ and $Y$.

Since the mutual information, defined in Eq. \ref{MIeq}, is symmetric, the network is undirected. Furthermore, the network is completely connected, because all the nodes are connected. However, the edges differ by their weights given by the mutual information.

\section{COMMUNITY DETECTION BY THE MULTILEVEL METHOD}
Proposed by Blondel et al. \cite{Blondel2008}, the MultiLevel algorithm (ML) is one of the community detection methods. Yang et al. \cite{Yang2016} compared several well-known algorithms of community detection (Edge-betweenness\cite{Edbet}, Fastgreedy \cite{fastg}, Infomap \cite{infom}, walktrap \cite{walkt}, and Spinglass \cite{SPG}), and found that ML outperforms all other algorithms on a set of benchmarks.

The ML algorithm aims to optimise the modularity \cite{Modu}, which measures the density of links inside a community, and compares it between other communities. The modularity is defined as follows:
\begin{footnotesize}
\begin{equation}
\label{Modu}
Q=\frac{1}{2m}\sum_{ij}\left[A_{ij}-\frac{k_ik_j}{2m}\delta(c_i,c_j)\right] 
\end{equation}
\end{footnotesize}
where $Q$ ranges between $-1$ and $1$ and\cite{Blondel2008}:
\begin{itemize}
\item $A_{ij}$ is the weight between nodes $i$ and $j$;
\item $2m$ is the sum of all the weights in the graph;
\item $k_i$ and $k_j$ are the sum of weights connected to nodes $i$ and $j$ respectively;
\item $c_i$ and $c_j$ communities (classes) of nodes.
\item $\delta$ is the delta function of the variables $c_i$ and $c_j$.
\end{itemize}

The ML algorithm consists of two iterative steps. Firstly, each node is considered as a community for an initial partition. Then, the node $i$ is removed from its community $c_i$ and placed in another community $c_j$, if this replacement maximises the modularity (Eq. \ref{Modu}), otherwise the node $i$ remains in its original community until when there is no gain in the modularity. The gain in modularity of moving a node $i$ into a community $C$ is computed as follows \cite{Blondel2008}:
\begin{footnotesize}
\begin{equation}
\label{Gmeq}
\Delta Q= \left[ \frac{\sum_{in}+2k_{i,in}}{2m}- \left( \frac{\sum_{tot}+k_i}{2m}\right)^2\right] - \left[ \frac{\sum_{in}}{2m}-\left( \frac{\sum_{tot}}{2m}\right)^2 - \left( \frac{k_i}{2m}\right)^2 \right] 
\end{equation}
\end{footnotesize}
where $\sum_{in}$ is the sum of weights inside $C$, $\sum_{tot}$  is the sum of weights of edges incident to nodes in community $C$, $k_{i,in}$ is the sum of weights of connection of node $i$ with other nodes of the same community $C$, and $m$ is the sum of all weights in the network.

In the second step, every community is considered as a node and building a new network. The weights between these new nodes are defined by the sum of the link weights of the corresponding communities of the old network, as it is proposed by Arenas et al. \cite{reduce} for reducing size of a complex network by preserving the modularity. Then, the first step is applied again on the new network iteratively until the modularity stops to increase.

\section{Results and discussion}
\begin{figure}
\centering
\includegraphics[width=\linewidth]{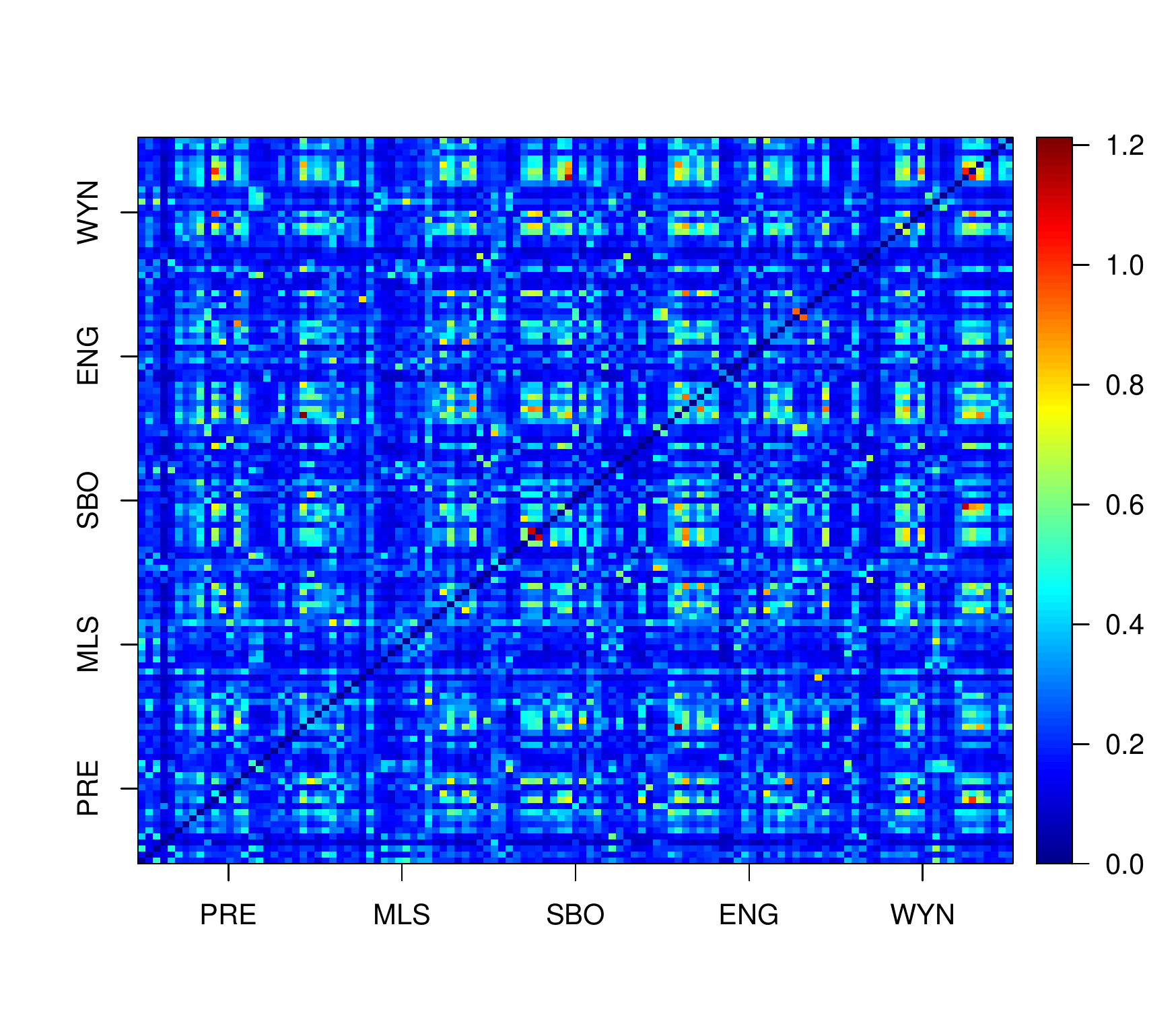}
\caption{Mutual information matrix $(119$ x $119)$. Each cell $(i,j)$ represents the mutual information between station $i$ and station $j$}
\label{fig3}  
\end{figure}

\begin{figure}
\centering
\includegraphics[width=\linewidth]{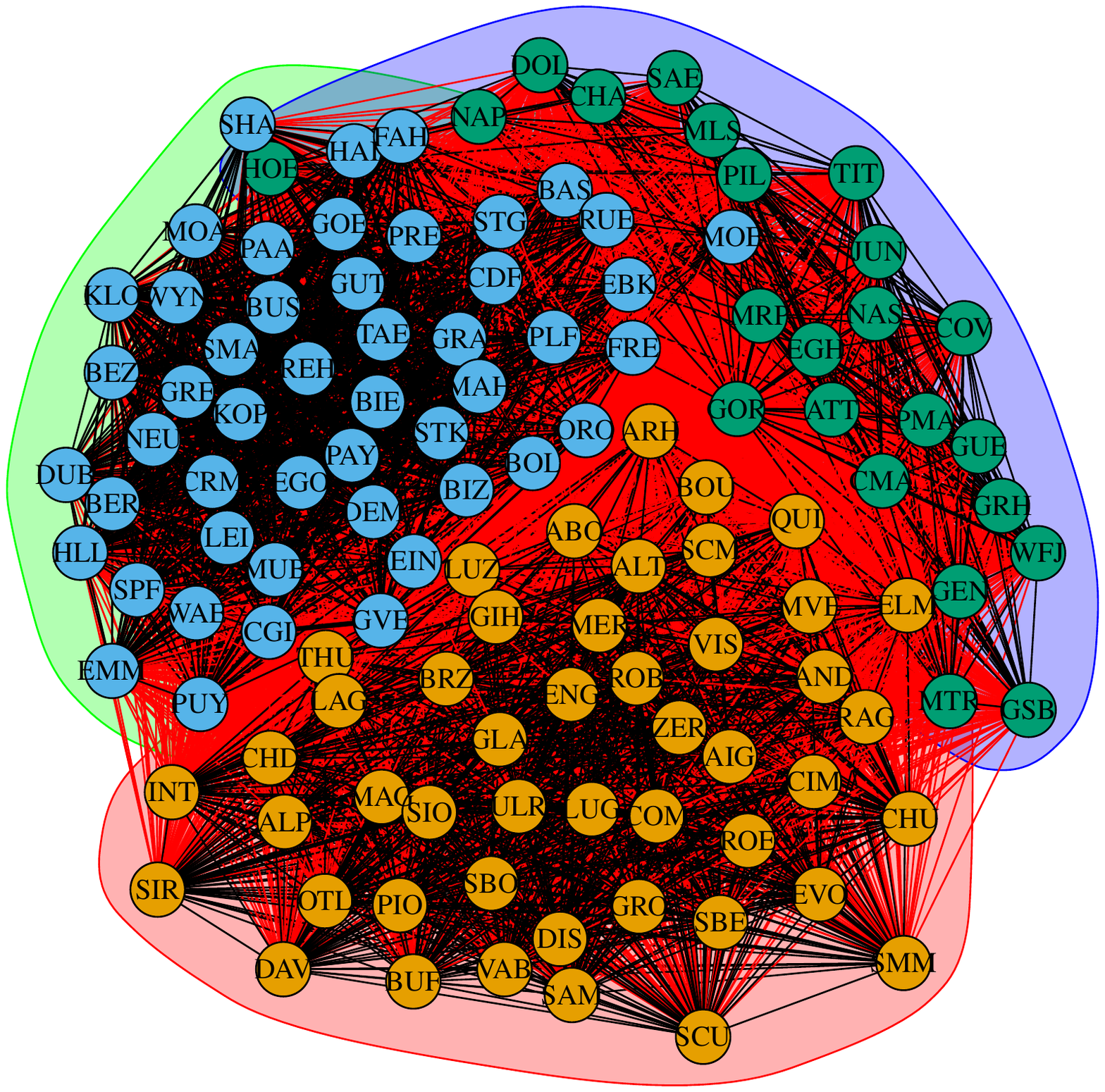}
\caption{Network visualisation with the three communities obtained by ML method before applying the STL.}
\label{fig4}  
\end{figure}

\begin{figure}
\centering
\includegraphics[width=\linewidth]{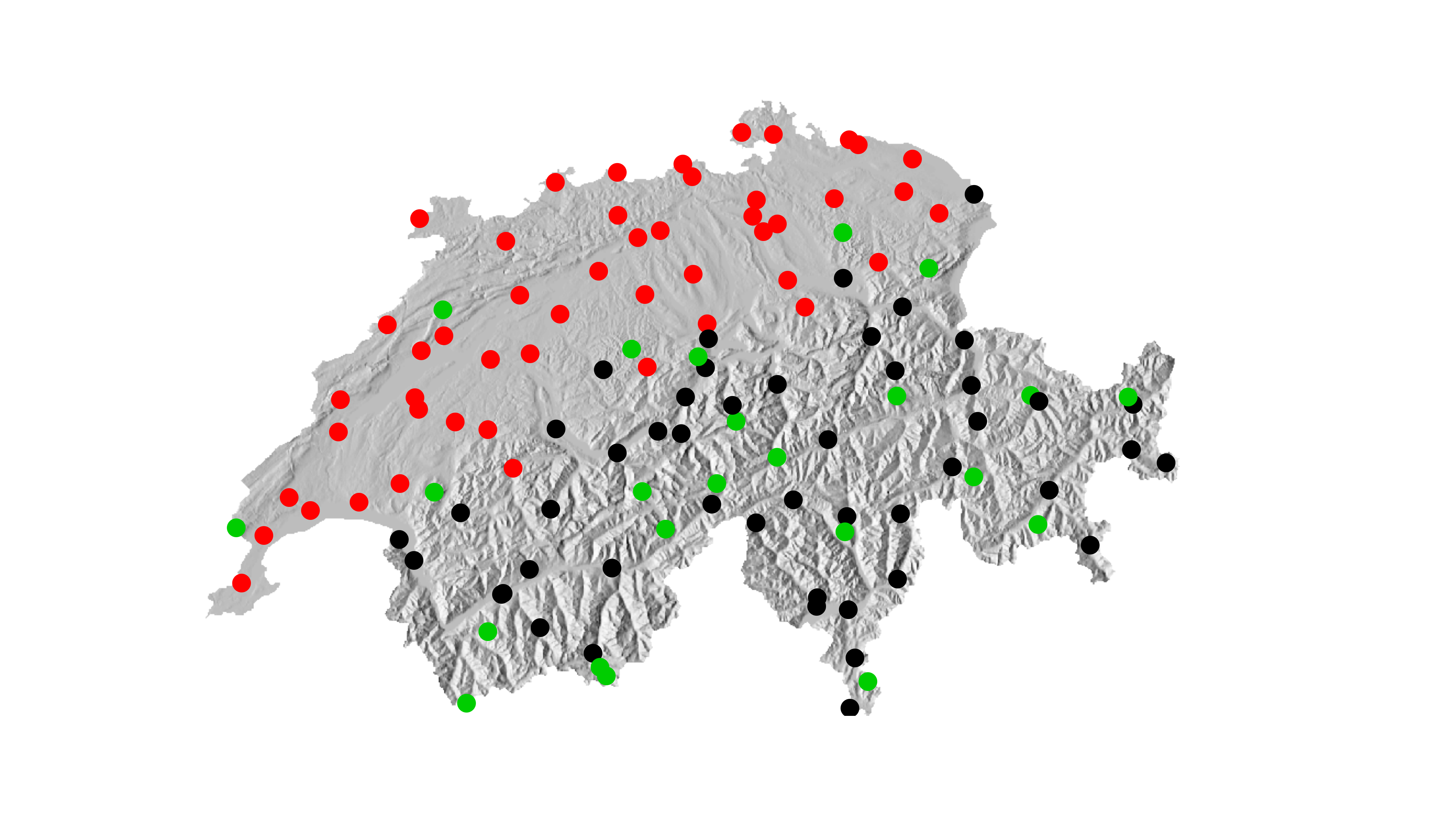}
\caption{Communities detected in network constructed before the STL decomposition.}
\label{fig5}  
\end{figure}

Fig. \ref{fig3} shows the mutual information among all the nodes.
Applying the community detection based on the MultiLevel method, three different communities are identified, as shown in Fig. \ref{fig4}. Mapping the communities on the territory of Switzerland (Fig. \ref{fig5}), two classes are mixed spatially (stations indicated by green and black circles).

To quantify such spatial mixing effect, the well-known silhouette width was used \cite{Silhouettes1987}.
This is defined as
\begin{equation}
\label{Sil}
s(i)=\frac{b(i)-a(i)}{max\{a(i),b(i)\}}
\end{equation}
where $a(i)$ is the dissimilarity between the node (object) $i$ and the other nodes of the same community, $b(i)$ is the minimum value of dissimilarity between the node $i$ and the other nodes of other communities, and the dissimilarity is the minimum Euclidean distance. From Eq. \ref{Sil}, we can see that the silhouette $s(i)$ ranges between $-1$ and $1$. 

\begin{figure}
\centering
\includegraphics[width=\linewidth]{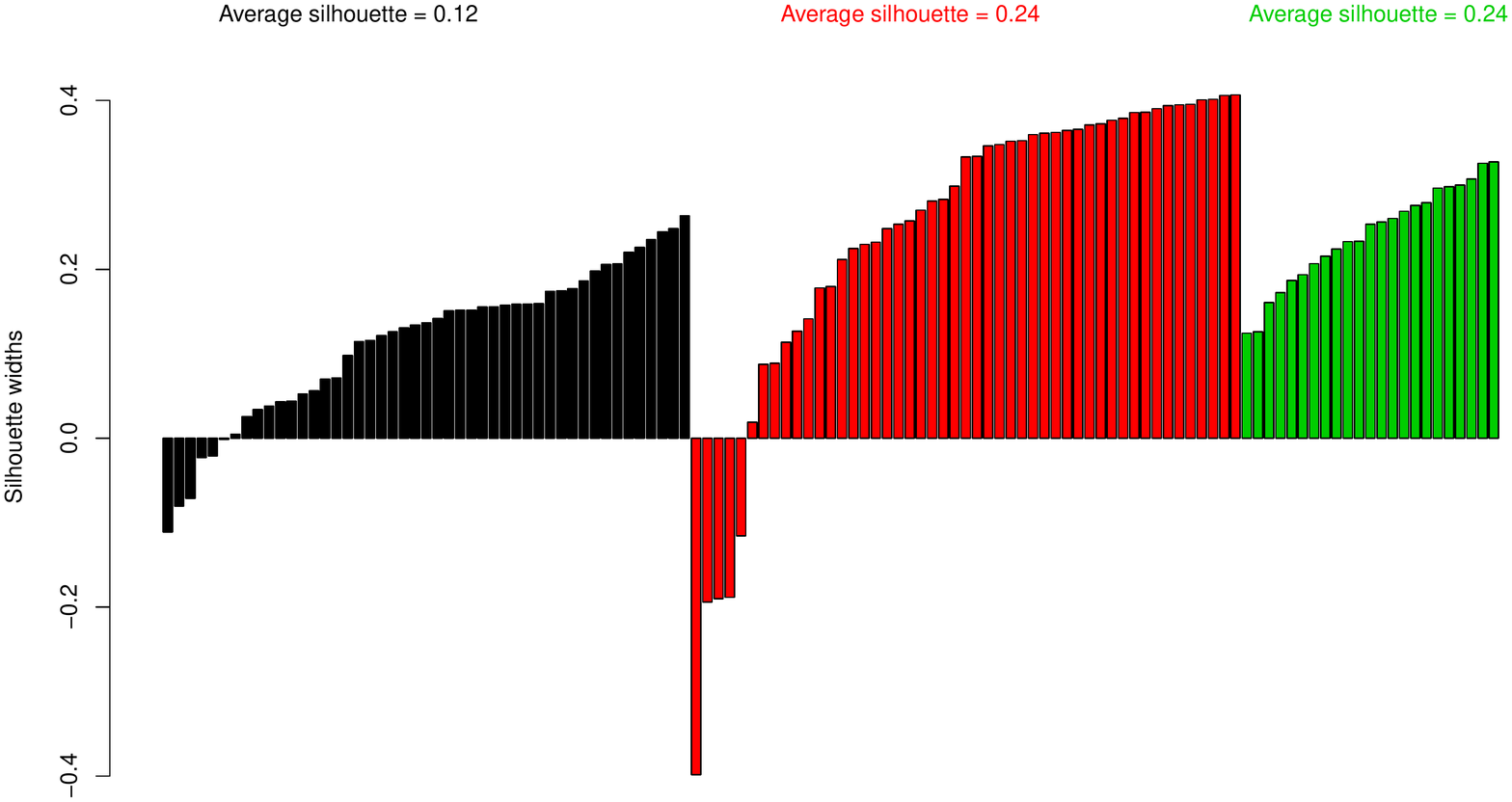}
\caption{Silhouette width of each node of each community obtained, on the Mutual information matrix, before applying the STL decomposition. The average value of the silhouette widths are: $0.12$ for the first community (black), $0.24$ for the second community (red), $0.24$ for the third community (green). The total average is $0.19$}
\label{fig6}  
\end{figure}

\begin{figure}
\centering
\includegraphics[width=\linewidth]{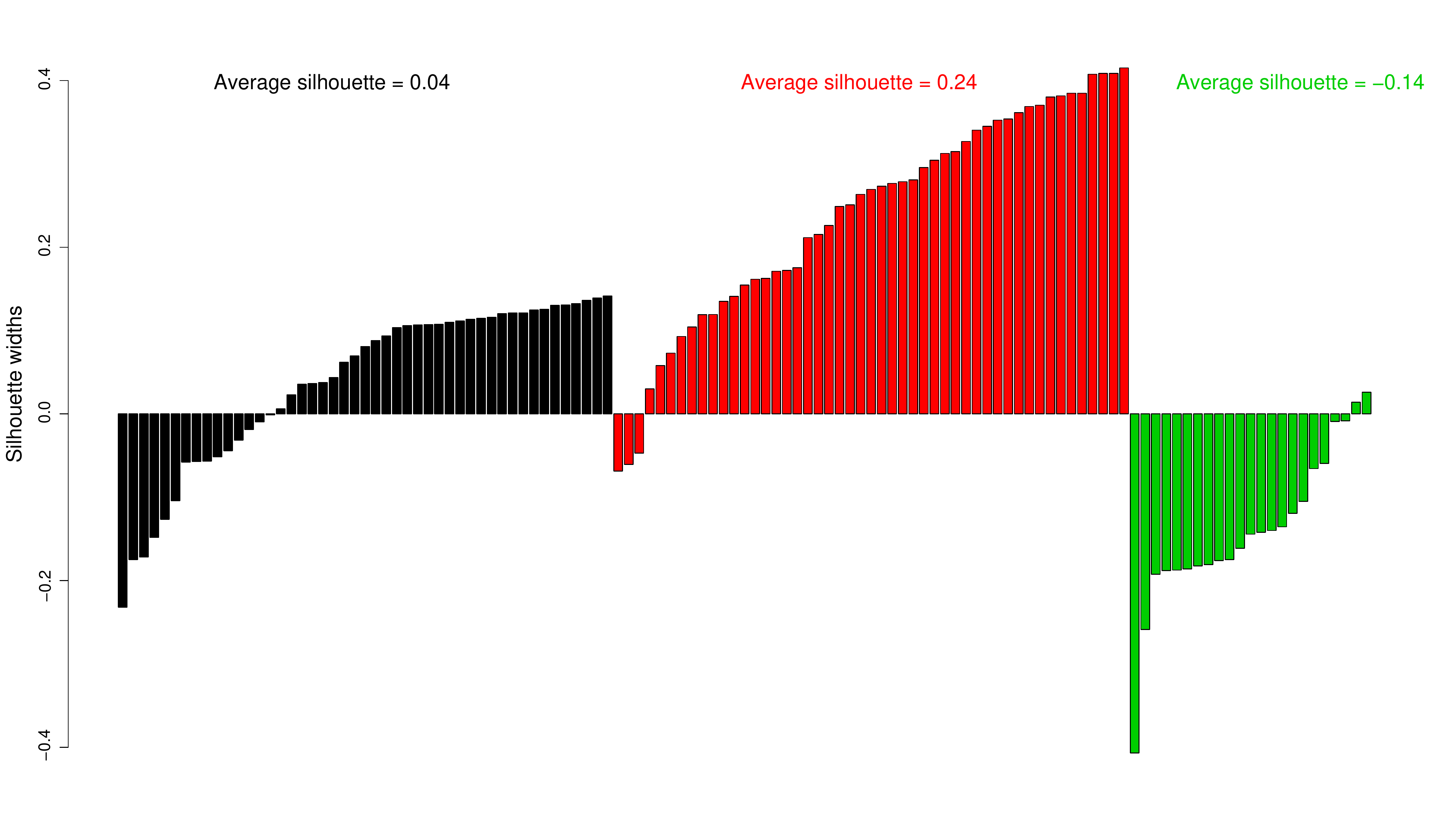}
\caption{Silhouette width of each node of each community obtained, on the XY coordinates, before applying the STL decomposition. The average value of the silhouette widths are: $0.04$ for the first community (black), $0.24$ for the second community (red), $-0.14$ for the third community (green). The total average is $0.09$ }
\label{fig7}  
\end{figure}

Fig. \ref{fig6} shows the silhouette widths for each station of each community, by applying the silhouette on the mutual information matrix and the obtained communities. Fig. \ref{fig7} shows the silhouette widths, by applying it on the XY coordinates.
The average values are $0.19$ (Mutual information matrix) and $0.09$ (XY coordinates). These low values indicate that the obtained communities are not well spatially separated.

\begin{figure}
\centering
\includegraphics[width=\linewidth]{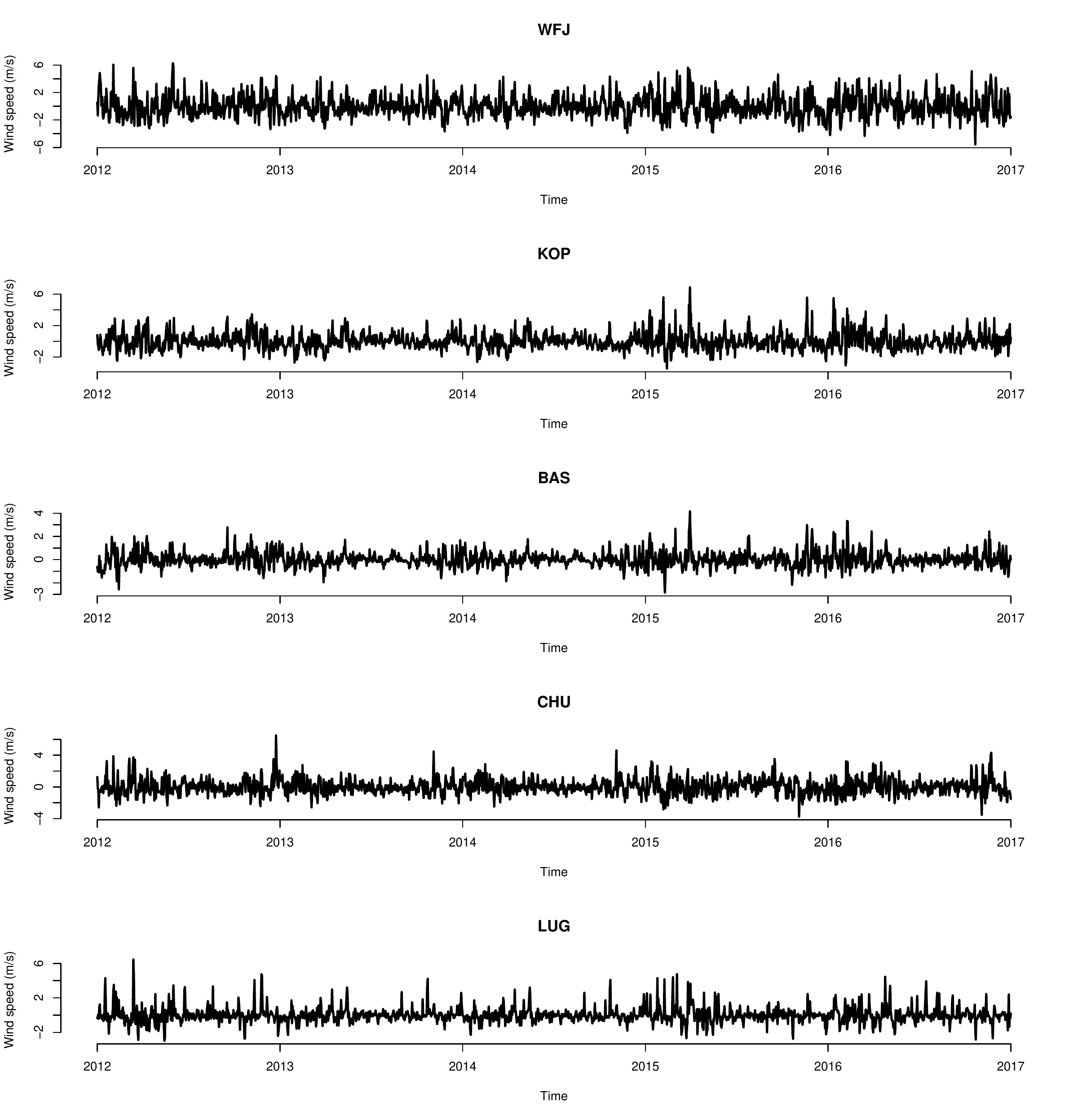}
\caption{Residuals of wind series, shown in Fig. \ref{fig2}, obtained by using the STL decomposition.}
\label{fig8}  
\end{figure}

\begin{figure}
\centering
\includegraphics[width=\linewidth]{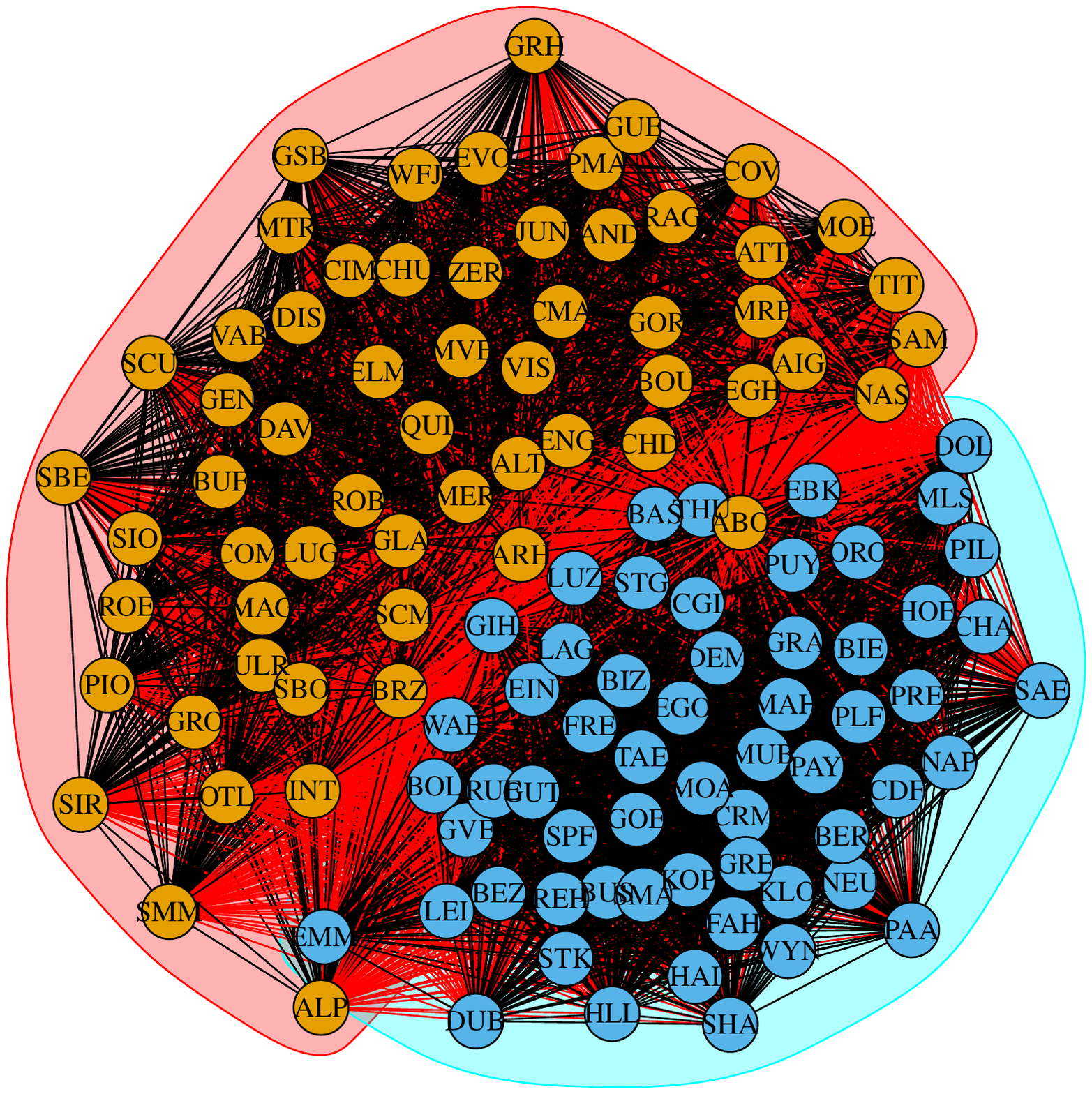}
\caption{Network visualisation with the two communities obtained by ML method after applying the STL.}
\label{fig9}  
\end{figure}

In order to understand the origin of such spatial mixing between communities, we filtered out from the wind series the trend and the yearly cycle \cite{Laib2018c} by using the Seasonal Decomposition of Time Series by Loess (STL) \cite{Cleveland1990} (implemented by using the stl function of the ”stats” R library \cite{lanR}). Then, we applied the community detection MultiLevel method to the residual wind series. Fig. \ref{fig8} shows the residuals of the same time series showed in Fig. \ref{fig2}, and Fig. \ref{fig9} presents two detected communities.

\begin{figure}
\centering
\includegraphics[width=\linewidth]{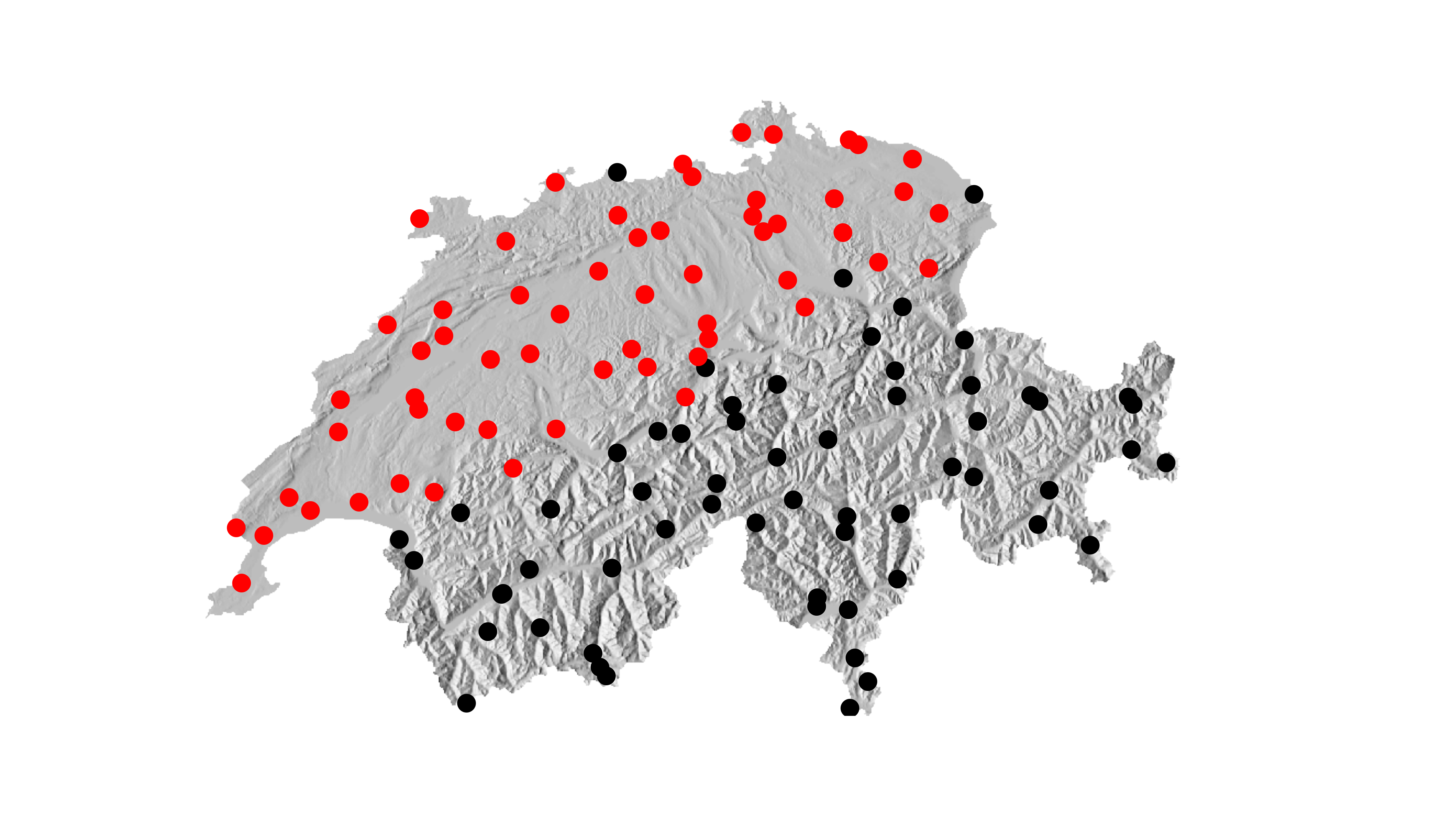}
\caption{Communities detected in network constructed after the STL decomposition.}
\label{fig10}  
\end{figure}

Mapping the communities on the Swiss territory, the two communities do not show significant spatial mixing (Fig. \ref{fig10}).

\begin{figure}
\centering
\includegraphics[width=\linewidth]{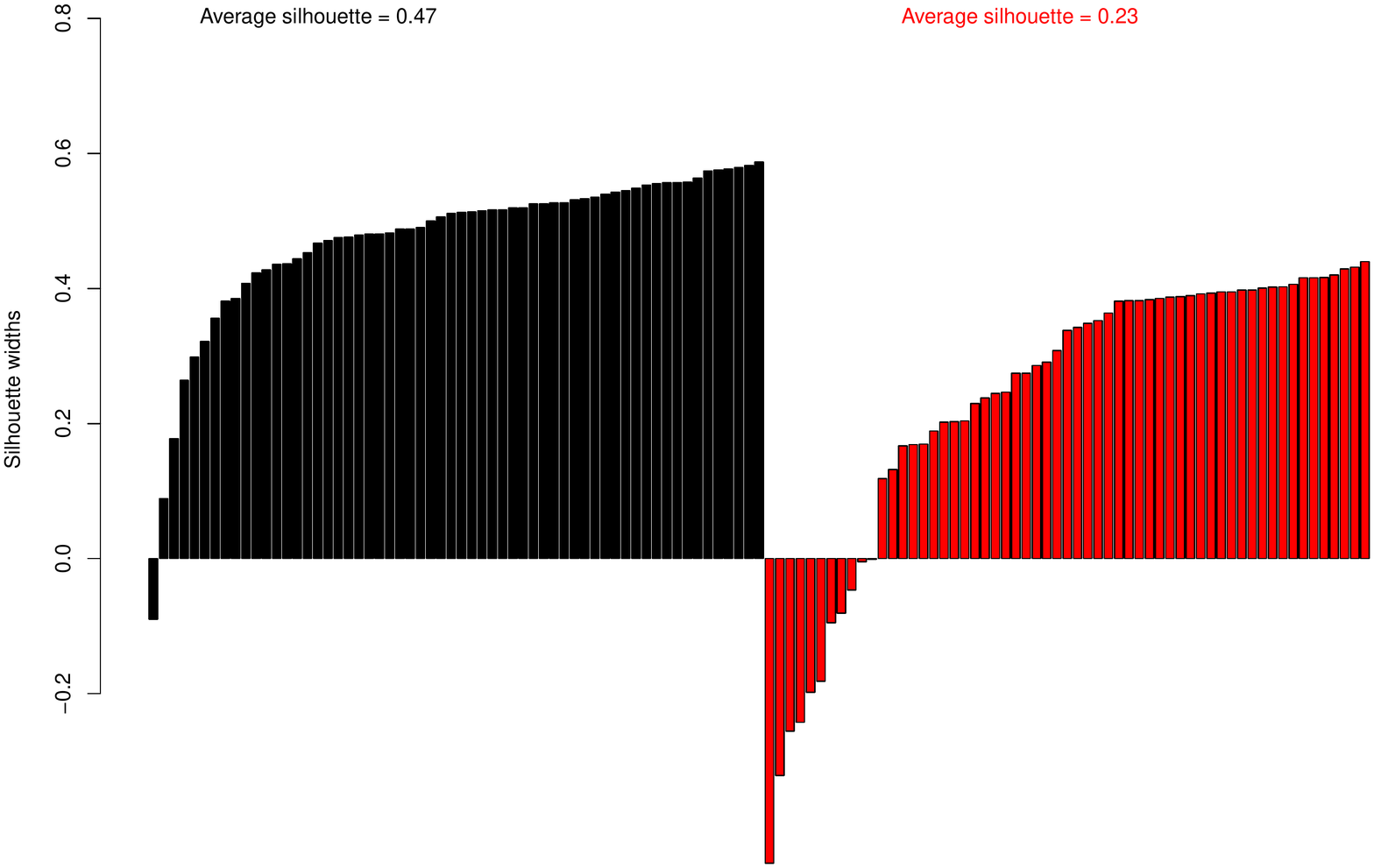}
\caption{Silhouette width of each node of each community obtained, on the mutual information matrix, after applying the STL decomposition. The average value of the silhouette widths are: $0.47$ for the first community (black), $0.23$ for the second community (red). The total average is $0.35$}
\label{fig11}  
\end{figure}

\begin{figure}
\centering
\includegraphics[width=\linewidth]{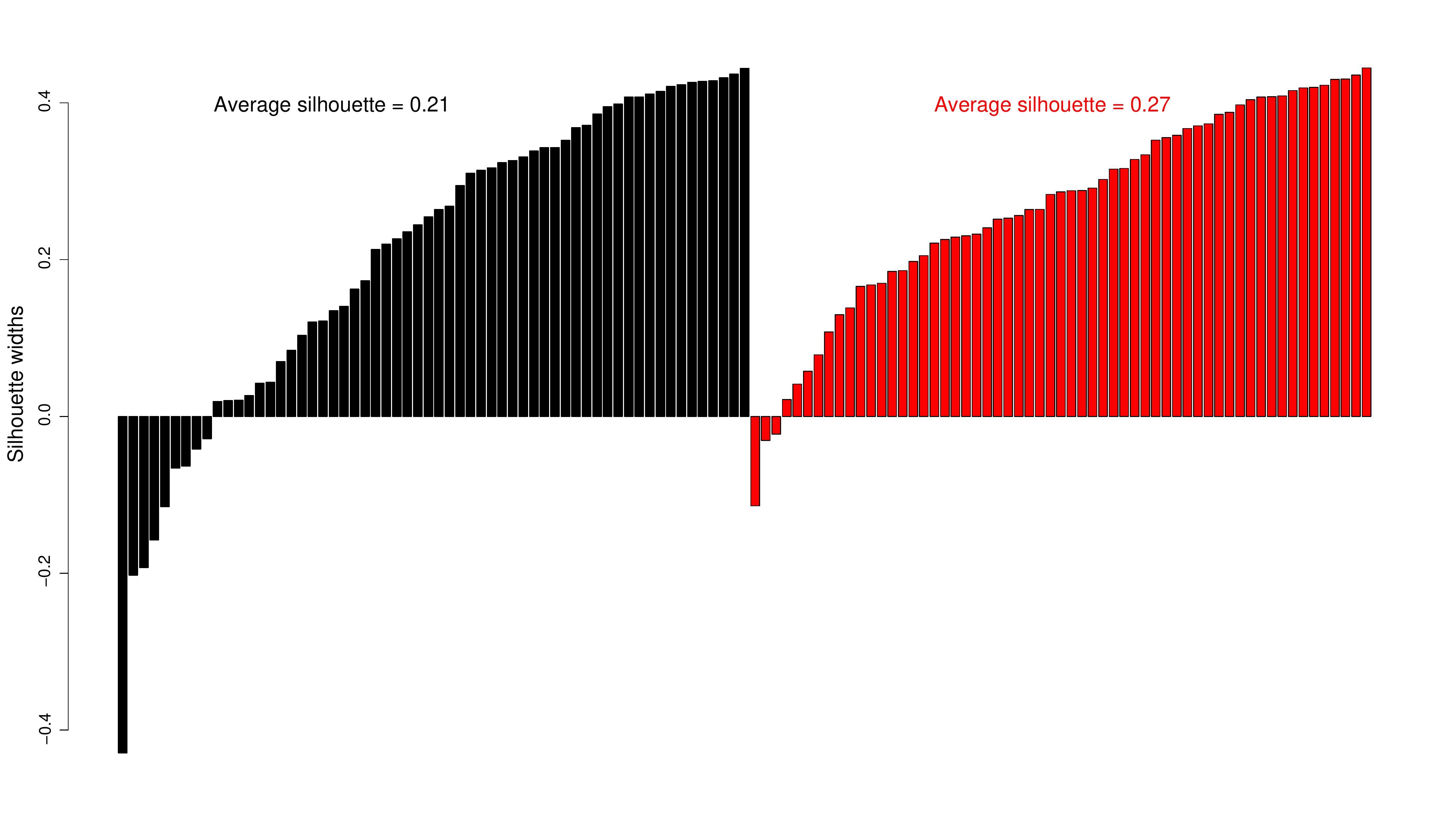}
\caption{Silhouette width of each node of each community obtained, on the XY coordinates, after applying the STL decomposition. The average value of the silhouette widths are: $0.21$ for the first community (black), $0.27$ for the second community (red). The total average is $0.24$}
\label{fig12}  
\end{figure}

\begin{figure}
\centering
\includegraphics[width=\linewidth]{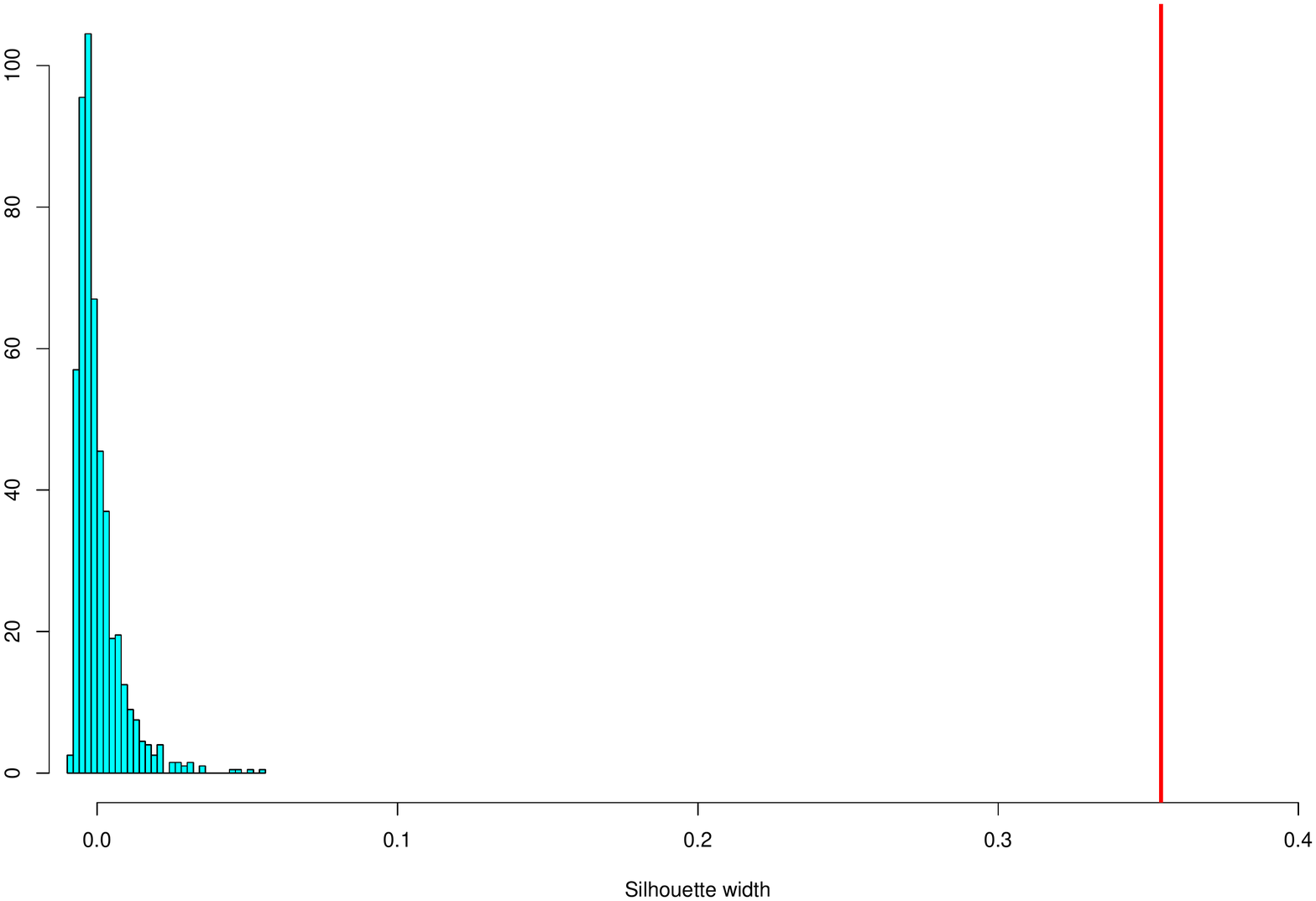}
\caption{Comparison between the silhouette width (obtained using the mutual information matrix) histogram of $1,000$ random classes (blue) and the total average silhouette width for classes obtained after STL decomposition (red).}
\label{fig13}  
\end{figure}

\begin{figure}
\centering
\includegraphics[width=\linewidth]{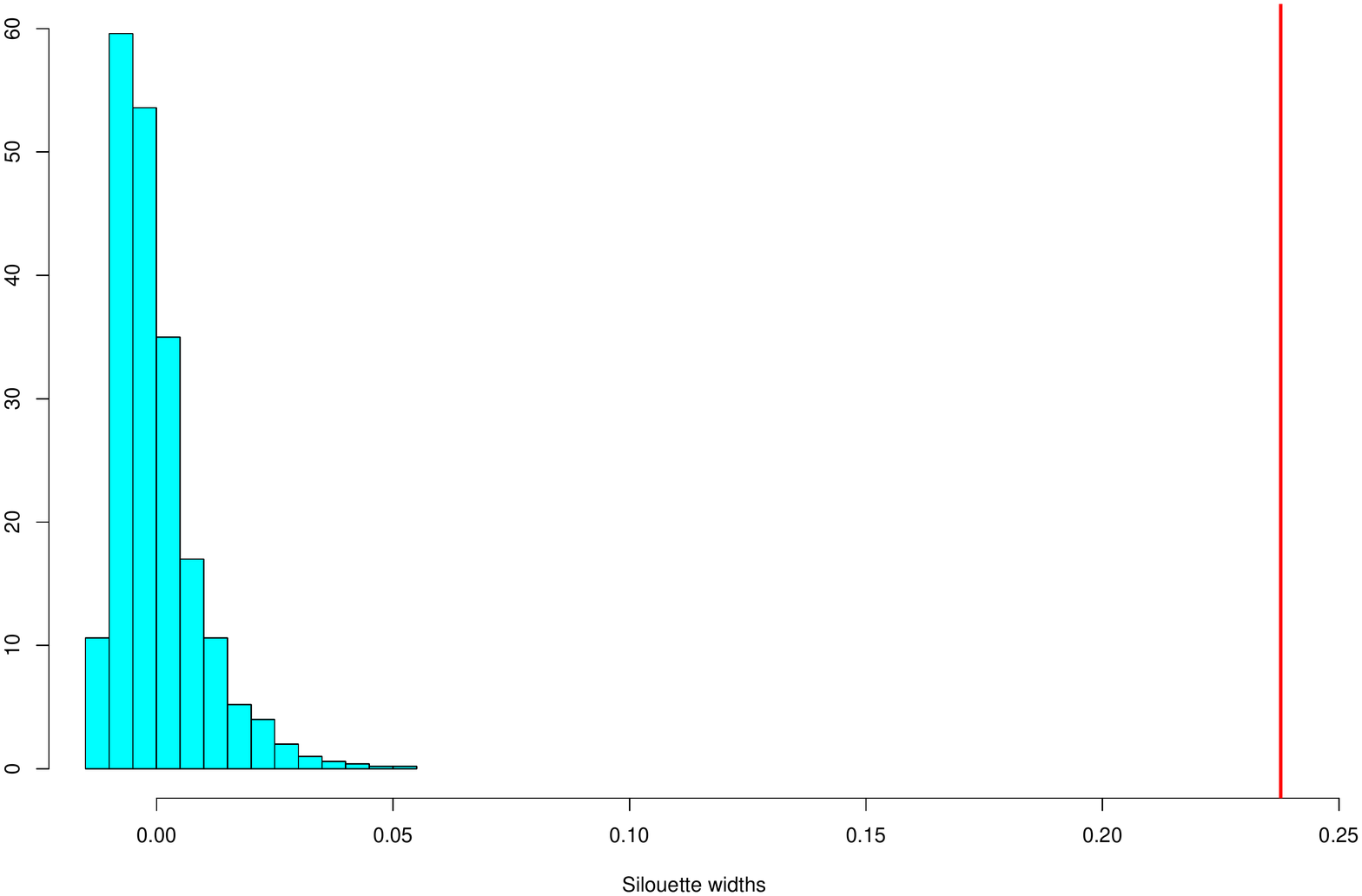}
\caption{Comparison between the silhouette width (obtained using the XY coordinates) histogram of $1,000$ random classes (blue) and the total average silhouette width for classes obtained after STL decomposition (red).}
\label{fig14}  
\end{figure}

Furthermore, the silhouette width for each station of each community is shown in Figs. \ref{fig11} and \ref{fig12}, and the mean silhouette values are $0.35$ (mutual information matrix) and $0.24$ (XY coordinates) respectively. These values are better than that obtained on the original data before applying STL. This indicates that there is not significant spatial mixing  between the two communities. This result was found also significant comparing it with the silhouette widths calculated for $1,000$ random spatial distribution of the stations. Figs. \ref{fig13} and \ref{fig14} show the histogram of the silhouette width for the randomised classes.

\section{Conclusions}
\begin{enumerate}

\item The wind network, constructed by representing the interactions between the nodes using the mutual information, highlights the (non-linear) correlations among the wind series.

\item The STL decomposition permits to extract the residuals of the wind speed not influenced by the trends and annual weather-induced forcings, but only by local meteo-climatic features depending on the geo-morphological and topographic characteristics of each measuring station.

\item The MultiLevel method for community detection in the mutual information-based network of wind series shows different topological structures of the monitoring system, before and after the removal of the trend and seasonal components. The network constructed on the original data is characterized by three different communities, while that constructed on the residual data (deprived of the trend and seasonal component) is characterized only by two communities.

\item The communities of the network built on the original data are quite spatially mixed. However, the communities of the network built on the residual data are, instead, spatially well separated, with no significantly apparent mixing between the stations belonging to the two communities.

\item The silhouette width, used to quantify the spatial mixing between the found communities, shows an average value for the communities detected in the network based on the original data much lower than that found for the communities detected in the network based on the residuals. Furthermore, the last is significant against the silhouette widths calculated after shuffling the stations of the two communities.

\item The two communities detected after removing the trend and seasonal components match very well with climatic zones of Switzerland, the Alps and the Jura-Plateau. This suggests the potential of the complex network method in disclosing the inner interactions among wind speed series measured in different climatic regions mainly due to the local topographic factors.

\end{enumerate}

\section{Acknowledgements}
F. Guignard thanks the support of the National Research Programme 75
"Big Data" (PNR75) of the Swiss National Science Foundation (SNSF).

 L. Telesca thanks the support of the "Scientific Exchanges" project n$^\circ$ 180296 funded by the SNSF.
 
M. Laib thanks the support of "Soci\'et\'e Acad\'emique Vaudoise" (SAV) and the Swiss Government Excellence Scholarships.

The authors thank MeteoSwiss for providing the data.


\bibliography{aipsamp}

\providecommand{\noopsort}[1]{}\providecommand{\singleletter}[1]{#1}%
\begin{thebibliography}{42}%
\makeatletter
\providecommand \@ifxundefined [1]{%
 \@ifx{#1\undefined}
}%
\providecommand \@ifnum [1]{%
 \ifnum #1\expandafter \@firstoftwo
 \else \expandafter \@secondoftwo
 \fi
}%
\providecommand \@ifx [1]{%
 \ifx #1\expandafter \@firstoftwo
 \else \expandafter \@secondoftwo
 \fi
}%
\providecommand \natexlab [1]{#1}%
\providecommand \enquote  [1]{``#1''}%
\providecommand \bibnamefont  [1]{#1}%
\providecommand \bibfnamefont [1]{#1}%
\providecommand \citenamefont [1]{#1}%
\providecommand \href@noop [0]{\@secondoftwo}%
\providecommand \href [0]{\begingroup \@sanitize@url \@href}%
\providecommand \@href[1]{\@@startlink{#1}\@@href}%
\providecommand \@@href[1]{\endgroup#1\@@endlink}%
\providecommand \@sanitize@url [0]{\catcode `\\12\catcode `\$12\catcode
  `\&12\catcode `\#12\catcode `\^12\catcode `\_12\catcode `\%12\relax}%
\providecommand \@@startlink[1]{}%
\providecommand \@@endlink[0]{}%
\providecommand \url  [0]{\begingroup\@sanitize@url \@url }%
\providecommand \@url [1]{\endgroup\@href {#1}{\urlprefix }}%
\providecommand \urlprefix  [0]{URL }%
\providecommand \Eprint [0]{\href }%
\providecommand \doibase [0]{http://dx.doi.org/}%
\providecommand \selectlanguage [0]{\@gobble}%
\providecommand \bibinfo  [0]{\@secondoftwo}%
\providecommand \bibfield  [0]{\@secondoftwo}%
\providecommand \translation [1]{[#1]}%
\providecommand \BibitemOpen [0]{}%
\providecommand \bibitemStop [0]{}%
\providecommand \bibitemNoStop [0]{.\EOS\space}%
\providecommand \EOS [0]{\spacefactor3000\relax}%
\providecommand \BibitemShut  [1]{\csname bibitem#1\endcsname}%
\let\auto@bib@innerbib\@empty
\bibitem [{\citenamefont {Arenas}\ \emph {et~al.}()\citenamefont {Arenas},
  \citenamefont {Duch}, \citenamefont {Fernandez},\ and\ \citenamefont
  {Gomez}}]{reduce}%
  \BibitemOpen
  \bibfield  {author} {\bibinfo {author} {\bibnamefont {Arenas}, \bibfnamefont
  {A.}}, \bibinfo {author} {\bibnamefont {Duch}, \bibfnamefont {J.}}, \bibinfo
  {author} {\bibnamefont {Fernandez}, \bibfnamefont {A.}}, \ and\ \bibinfo
  {author} {\bibnamefont {Gomez}, \bibfnamefont {S.}},\ }\bibfield  {title}
  {\enquote {\bibinfo {title} {Size reduction of complex networks preserving
  modularity},}\ }\href {\doibase 10.1088/1367-2630/9/6/176} {\bibfield
  {journal} {\bibinfo  {journal} {New Journal of Physics}\ }\textbf {\bibinfo
  {volume} {9}},\ \bibinfo {pages} {176}}\BibitemShut {NoStop}%
\bibitem [{\citenamefont {Blondel}\ \emph {et~al.}(2008)\citenamefont
  {Blondel}, \citenamefont {Guillaume}, \citenamefont {Lambiotte},\ and\
  \citenamefont {Lefebvre}}]{Blondel2008}%
  \BibitemOpen
  \bibfield  {author} {\bibinfo {author} {\bibnamefont {Blondel}, \bibfnamefont
  {V.~D.}}, \bibinfo {author} {\bibnamefont {Guillaume}, \bibfnamefont
  {J.-L.}}, \bibinfo {author} {\bibnamefont {Lambiotte}, \bibfnamefont {R.}}, \
  and\ \bibinfo {author} {\bibnamefont {Lefebvre}, \bibfnamefont {E.}},\
  }\bibfield  {title} {\enquote {\bibinfo {title} {Fast unfolding of
  communities in large networks},}\ }\href {\doibase
  10.1088/1742-5468/2008/10/P10008} {\bibfield  {journal} {\bibinfo  {journal}
  {Journal of Statistical Mechanics: Theory and Experiment}\ ,\ \bibinfo
  {pages} {P10008}} (\bibinfo {year} {2008})}\BibitemShut {NoStop}%
\bibitem [{\citenamefont {Boccaletti}\ \emph {et~al.}(2006)\citenamefont
  {Boccaletti}, \citenamefont {Latora}, \citenamefont {Moreno}, \citenamefont
  {Chavez},\ and\ \citenamefont {Hwang}}]{Boccaletti2006}%
  \BibitemOpen
  \bibfield  {author} {\bibinfo {author} {\bibnamefont {Boccaletti},
  \bibfnamefont {S.}}, \bibinfo {author} {\bibnamefont {Latora}, \bibfnamefont
  {V.}}, \bibinfo {author} {\bibnamefont {Moreno}, \bibfnamefont {Y.}},
  \bibinfo {author} {\bibnamefont {Chavez}, \bibfnamefont {M.}}, \ and\
  \bibinfo {author} {\bibnamefont {Hwang}, \bibfnamefont {D.-U.}},\ }\bibfield
  {title} {\enquote {\bibinfo {title} {Complex networks: Structure and
  dynamics},}\ }\href {\doibase 10.1016/j.physrep.2005.10.009} {\bibfield
  {journal} {\bibinfo  {journal} {Physics Reports}\ }\textbf {\bibinfo {volume}
  {424}},\ \bibinfo {pages} {175 -- 308} (\bibinfo {year} {2006})}\BibitemShut
  {NoStop}%
\bibitem [{\citenamefont {Cao}, \citenamefont {Zhang},\ and\ \citenamefont
  {Li}(2017)}]{Cao2017}%
  \BibitemOpen
  \bibfield  {author} {\bibinfo {author} {\bibnamefont {Cao}, \bibfnamefont
  {G.}}, \bibinfo {author} {\bibnamefont {Zhang}, \bibfnamefont {Q.}}, \ and\
  \bibinfo {author} {\bibnamefont {Li}, \bibfnamefont {Q.}},\ }\bibfield
  {title} {\enquote {\bibinfo {title} {Causal relationship between the global
  foreign exchange market based on complex networks and entropy theory},}\
  }\href {\doibase https://doi.org/10.1016/j.chaos.2017.03.039} {\bibfield
  {journal} {\bibinfo  {journal} {Chaos, Solitons \& Fractals}\ }\textbf
  {\bibinfo {volume} {99}},\ \bibinfo {pages} {36 -- 44} (\bibinfo {year}
  {2017})}\BibitemShut {NoStop}%
\bibitem [{\citenamefont {Clauset}, \citenamefont {Newman},\ and\ \citenamefont
  {Moore}(2004)}]{fastg}%
  \BibitemOpen
  \bibfield  {author} {\bibinfo {author} {\bibnamefont {Clauset}, \bibfnamefont
  {A.}}, \bibinfo {author} {\bibnamefont {Newman}, \bibfnamefont {M.~E.~J.}}, \
  and\ \bibinfo {author} {\bibnamefont {Moore}, \bibfnamefont {C.}},\
  }\bibfield  {title} {\enquote {\bibinfo {title} {Finding community structure
  in very large networks},}\ }\href {\doibase 10.1103/PhysRevE.70.066111}
  {\bibfield  {journal} {\bibinfo  {journal} {Phys. Rev. E}\ }\textbf {\bibinfo
  {volume} {70}},\ \bibinfo {pages} {066111} (\bibinfo {year}
  {2004})}\BibitemShut {NoStop}%
\bibitem [{\citenamefont {Cleveland}\ \emph {et~al.}(1990)\citenamefont
  {Cleveland}, \citenamefont {Cleveland}, \citenamefont {McRae},\ and\
  \citenamefont {Terpenning}}]{Cleveland1990}%
  \BibitemOpen
  \bibfield  {author} {\bibinfo {author} {\bibnamefont {Cleveland},
  \bibfnamefont {R.~B.}}, \bibinfo {author} {\bibnamefont {Cleveland},
  \bibfnamefont {W.~S.}}, \bibinfo {author} {\bibnamefont {McRae},
  \bibfnamefont {J.~E.}}, \ and\ \bibinfo {author} {\bibnamefont {Terpenning},
  \bibfnamefont {I.}},\ }\bibfield  {title} {\enquote {\bibinfo {title} {Stl: A
  seasonal-trend decomposition procedure based on loess},}\ }\href@noop {}
  {\bibfield  {journal} {\bibinfo  {journal} {Journal of Official Statistics}\
  }\textbf {\bibinfo {volume} {6}},\ \bibinfo {pages} {3 -- 73} (\bibinfo
  {year} {1990})}\BibitemShut {NoStop}%
\bibitem [{\citenamefont {COVER}\ and\ \citenamefont {THOMAS}(2006)}]{MIbook}%
  \BibitemOpen
  \bibfield  {author} {\bibinfo {author} {\bibnamefont {COVER}, \bibfnamefont
  {T.~M.}}\ and\ \bibinfo {author} {\bibnamefont {THOMAS}, \bibfnamefont
  {J.~A.}},\ }\href@noop {} {\emph {\bibinfo {title} {ELEMENTS OF INFORMATION
  THEORY. 2nd ed.}}}\ (\bibinfo  {publisher} {Wiley-interscience},\ \bibinfo
  {year} {2006})\ p.\ \bibinfo {pages} {774}\BibitemShut {NoStop}%
\bibitem [{\citenamefont {D'Amico}, \citenamefont {Petroni},\ and\
  \citenamefont {Prattico}(2015)}]{DAMICO2015}%
  \BibitemOpen
  \bibfield  {author} {\bibinfo {author} {\bibnamefont {D'Amico}, \bibfnamefont
  {G.}}, \bibinfo {author} {\bibnamefont {Petroni}, \bibfnamefont {F.}}, \ and\
  \bibinfo {author} {\bibnamefont {Prattico}, \bibfnamefont {F.}},\ }\bibfield
  {title} {\enquote {\bibinfo {title} {Wind speed prediction for wind farm
  applications by extreme value theory and copulas},}\ }\href@noop {}
  {\bibfield  {journal} {\bibinfo  {journal} {\mbox{Wind Eng. Ind. Aerodyn. }
  Journal}\ }\textbf {\bibinfo {volume} {145}},\ \bibinfo {pages} {p. 229--236}
  (\bibinfo {year} {2015})}\BibitemShut {NoStop}%
\bibitem [{\citenamefont {Donges}\ \emph
  {et~al.}(2009{\natexlab{a}})\citenamefont {Donges}, \citenamefont {Zou},
  \citenamefont {Marwan},\ and\ \citenamefont {Kurths}}]{Donges2009a}%
  \BibitemOpen
  \bibfield  {author} {\bibinfo {author} {\bibnamefont {Donges}, \bibfnamefont
  {J.~F.}}, \bibinfo {author} {\bibnamefont {Zou}, \bibfnamefont {Y.}},
  \bibinfo {author} {\bibnamefont {Marwan}, \bibfnamefont {N.}}, \ and\
  \bibinfo {author} {\bibnamefont {Kurths}, \bibfnamefont {J.}},\ }\bibfield
  {title} {\enquote {\bibinfo {title} {The backbone of the climate network},}\
  }\href@noop {} {\bibfield  {journal} {\bibinfo  {journal} {EPL (Europhysics
  Letters)}\ }\textbf {\bibinfo {volume} {87}},\ \bibinfo {pages} {48007}
  (\bibinfo {year} {2009}{\natexlab{a}})}\BibitemShut {NoStop}%
\bibitem [{\citenamefont {Donges}\ \emph
  {et~al.}(2009{\natexlab{b}})\citenamefont {Donges}, \citenamefont {Zou},
  \citenamefont {Marwan},\ and\ \citenamefont {Kurths}}]{Donges2009b}%
  \BibitemOpen
  \bibfield  {author} {\bibinfo {author} {\bibnamefont {Donges}, \bibfnamefont
  {J.~F.}}, \bibinfo {author} {\bibnamefont {Zou}, \bibfnamefont {Y.}},
  \bibinfo {author} {\bibnamefont {Marwan}, \bibfnamefont {N.}}, \ and\
  \bibinfo {author} {\bibnamefont {Kurths}, \bibfnamefont {J.}},\ }\bibfield
  {title} {\enquote {\bibinfo {title} {Complex networks in climate dynamics},}\
  }\href@noop {} {\bibfield  {journal} {\bibinfo  {journal} {The European
  Physical Journal Special Topics}\ }\textbf {\bibinfo {volume} {174}},\
  \bibinfo {pages} {157--179} (\bibinfo {year}
  {2009}{\natexlab{b}})}\BibitemShut {NoStop}%
\bibitem [{\citenamefont {Donner}\ \emph {et~al.}(2019)\citenamefont {Donner},
  \citenamefont {Lindner}, \citenamefont {Tupikina},\ and\ \citenamefont
  {Molkenthin}}]{Donner2019}%
  \BibitemOpen
  \bibfield  {author} {\bibinfo {author} {\bibnamefont {Donner}, \bibfnamefont
  {R.~V.}}, \bibinfo {author} {\bibnamefont {Lindner}, \bibfnamefont {M.}},
  \bibinfo {author} {\bibnamefont {Tupikina}, \bibfnamefont {L.}}, \ and\
  \bibinfo {author} {\bibnamefont {Molkenthin}, \bibfnamefont {N.}},\ }\enquote
  {\bibinfo {title} {Characterizing flows by complex network methods},}\ in\
  \href {\doibase 10.1007/978-3-319-78512-7_11} {\emph {\bibinfo {booktitle} {A
  Mathematical Modeling Approach from Nonlinear Dynamics to Complex
  Systems}}},\ \bibinfo {editor} {edited by\ \bibinfo {editor} {\bibfnamefont
  {E.~E.~N.}\ \bibnamefont {Macau}}}\ (\bibinfo  {publisher} {Springer
  International Publishing},\ \bibinfo {address} {Cham},\ \bibinfo {year}
  {2019})\ pp.\ \bibinfo {pages} {197--226}\BibitemShut {NoStop}%
\bibitem [{\citenamefont {Donner}, \citenamefont {Wiedermann},\ and\
  \citenamefont {Donges}(2017)}]{Donner2}%
  \BibitemOpen
  \bibfield  {author} {\bibinfo {author} {\bibnamefont {Donner}, \bibfnamefont
  {R.~V.}}, \bibinfo {author} {\bibnamefont {Wiedermann}, \bibfnamefont {M.}},
  \ and\ \bibinfo {author} {\bibnamefont {Donges}, \bibfnamefont {J.~F.}},\
  }\enquote {\bibinfo {title} {Complex network techniques for climatological
  data analysis},}\ in\ \href {\doibase 10.1017/9781316339251.007} {\emph
  {\bibinfo {booktitle} {Nonlinear and Stochastic Climate Dynamics}}},\
  \bibinfo {editor} {edited by\ \bibinfo {editor} {\bibfnamefont {C.~L.~E.}\
  \bibnamefont {Franzke}}\ and\ \bibinfo {editor} {\bibfnamefont {T.~J.}\
  \bibnamefont {O'Kane}}}\ (\bibinfo  {publisher} {Cambridge University
  Press},\ \bibinfo {year} {2017})\ p.\ \bibinfo {pages}
  {159–183}\BibitemShut {NoStop}%
\bibitem [{\citenamefont {Fortuna}, \citenamefont {Nunnari},\ and\
  \citenamefont {Guariso}(2014)}]{Fortuna2014}%
  \BibitemOpen
  \bibfield  {author} {\bibinfo {author} {\bibnamefont {Fortuna}, \bibfnamefont
  {L.}}, \bibinfo {author} {\bibnamefont {Nunnari}, \bibfnamefont {S.}}, \ and\
  \bibinfo {author} {\bibnamefont {Guariso}, \bibfnamefont {G.}},\ }\bibfield
  {title} {\enquote {\bibinfo {title} {Fractal order evidences in wind speed
  time series},}\ }\href@noop {} {\bibfield  {journal} {\bibinfo  {journal}
  {ICFDA'14 International Conference on Fractional Differentiation and Its
  Applications 2014}\ ,\ \bibinfo {pages} {1--6}} (\bibinfo {year}
  {2014})}\BibitemShut {NoStop}%
\bibitem [{\citenamefont {Fortunato}(2010)}]{Fortunato2010}%
  \BibitemOpen
  \bibfield  {author} {\bibinfo {author} {\bibnamefont {Fortunato},
  \bibfnamefont {S.}},\ }\bibfield  {title} {\enquote {\bibinfo {title}
  {Community detection in graphs},}\ }\href {\doibase
  10.1016/j.physrep.2009.11.002} {\bibfield  {journal} {\bibinfo  {journal}
  {Physics Reports}\ }\textbf {\bibinfo {volume} {486}},\ \bibinfo {pages} {75
  -- 174} (\bibinfo {year} {2010})}\BibitemShut {NoStop}%
\bibitem [{\citenamefont {Garcia-Marin}\ \emph {et~al.}(2013)\citenamefont
  {Garcia-Marin}, \citenamefont {Estévez}, \citenamefont {Jim\'enez-Hornero},\
  and\ \citenamefont {Ayuso-Munoz}}]{Garcia2013}%
  \BibitemOpen
  \bibfield  {author} {\bibinfo {author} {\bibnamefont {Garcia-Marin},
  \bibfnamefont {A.~P.}}, \bibinfo {author} {\bibnamefont {Estévez},
  \bibfnamefont {J.}}, \bibinfo {author} {\bibnamefont {Jim\'enez-Hornero},
  \bibfnamefont {F.~J.}}, \ and\ \bibinfo {author} {\bibnamefont {Ayuso-Munoz},
  \bibfnamefont {J.~L.}},\ }\bibfield  {title} {\enquote {\bibinfo {title}
  {Multifractal analysis of validated wind speed time series},}\ }\href
  {\doibase 10.1063/1.4793781} {\bibfield  {journal} {\bibinfo  {journal}
  {Chaos: An Interdisciplinary Journal of Nonlinear Science}\ }\textbf
  {\bibinfo {volume} {23}},\ \bibinfo {pages} {013133} (\bibinfo {year}
  {2013})}\BibitemShut {NoStop}%
\bibitem [{\citenamefont {Girvan}\ and\ \citenamefont
  {Newman}(2002{\natexlab{a}})}]{Girvan2002}%
  \BibitemOpen
  \bibfield  {author} {\bibinfo {author} {\bibnamefont {Girvan}, \bibfnamefont
  {M.}}\ and\ \bibinfo {author} {\bibnamefont {Newman}, \bibfnamefont
  {M.~E.~J.}},\ }\bibfield  {title} {\enquote {\bibinfo {title} {Community
  structure in social and biological networks},}\ }\href {\doibase
  10.1073/pnas.122653799} {\bibfield  {journal} {\bibinfo  {journal}
  {Proceedings of the National Academy of Sciences}\ }\textbf {\bibinfo
  {volume} {99}},\ \bibinfo {pages} {7821--7826} (\bibinfo {year}
  {2002}{\natexlab{a}})}\BibitemShut {NoStop}%
\bibitem [{\citenamefont {Girvan}\ and\ \citenamefont
  {Newman}(2002{\natexlab{b}})}]{Edbet}%
  \BibitemOpen
  \bibfield  {author} {\bibinfo {author} {\bibnamefont {Girvan}, \bibfnamefont
  {M.}}\ and\ \bibinfo {author} {\bibnamefont {Newman}, \bibfnamefont
  {M.~E.~J.}},\ }\bibfield  {title} {\enquote {\bibinfo {title} {Community
  structure in social and biological networks},}\ }\href {\doibase
  10.1073/pnas.122653799} {\bibfield  {journal} {\bibinfo  {journal}
  {Proceedings of the National Academy of Sciences}\ }\textbf {\bibinfo
  {volume} {99}},\ \bibinfo {pages} {7821--7826} (\bibinfo {year}
  {2002}{\natexlab{b}})}\BibitemShut {NoStop}%
\bibitem [{\citenamefont {Gozolchiani}\ \emph {et~al.}(2008)\citenamefont
  {Gozolchiani}, \citenamefont {Yamasaki}, \citenamefont {Gazit},\ and\
  \citenamefont {Havlin}}]{Gozolchiani2008}%
  \BibitemOpen
  \bibfield  {author} {\bibinfo {author} {\bibnamefont {Gozolchiani},
  \bibfnamefont {A.}}, \bibinfo {author} {\bibnamefont {Yamasaki},
  \bibfnamefont {K.}}, \bibinfo {author} {\bibnamefont {Gazit}, \bibfnamefont
  {O.}}, \ and\ \bibinfo {author} {\bibnamefont {Havlin}, \bibfnamefont {S.}},\
  }\bibfield  {title} {\enquote {\bibinfo {title} {Pattern of climate network
  blinking links follows el {N}iño events},}\ }\href@noop {} {\bibfield
  {journal} {\bibinfo  {journal} {EPL (Europhysics Letters)}\ }\textbf
  {\bibinfo {volume} {83}},\ \bibinfo {pages} {28005} (\bibinfo {year}
  {2008})}\BibitemShut {NoStop}%
\bibitem [{\citenamefont {Holger~Kantz}, \citenamefont {Ragwitz},\ and\
  \citenamefont {Vitanov}(2004)}]{KANTZ2004}%
  \BibitemOpen
  \bibfield  {author} {\bibinfo {author} {\bibnamefont {Holger~Kantz},
  \bibfnamefont {D.~H.}}, \bibinfo {author} {\bibnamefont {Ragwitz},
  \bibfnamefont {M.}}, \ and\ \bibinfo {author} {\bibnamefont {Vitanov},
  \bibfnamefont {N.~K.}},\ }\bibfield  {title} {\enquote {\bibinfo {title}
  {Markov chain model for turbulent wind speed data},}\ }\href@noop {}
  {\bibfield  {journal} {\bibinfo  {journal} {Physica A: Statistical Mechanics
  and its Applications}\ }\textbf {\bibinfo {volume} {342}},\ \bibinfo {pages}
  {315 -- 321} (\bibinfo {year} {2004})},\ \bibinfo {note} {proceedings of the
  VIII Latin American Workshop on Nonlinear Phenomena}\BibitemShut {NoStop}%
\bibitem [{\citenamefont {Jiménez}(2013)}]{Jimenez2013}%
  \BibitemOpen
  \bibfield  {author} {\bibinfo {author} {\bibnamefont {Jiménez},
  \bibfnamefont {A.}},\ }\bibfield  {title} {\enquote {\bibinfo {title} {A
  complex network model for seismicity based on mutual information},}\ }\href
  {\doibase https://doi.org/10.1016/j.physa.2013.01.062} {\bibfield  {journal}
  {\bibinfo  {journal} {Physica A: Statistical Mechanics and its Applications}\
  }\textbf {\bibinfo {volume} {392}},\ \bibinfo {pages} {2498 -- 2506}
  (\bibinfo {year} {2013})}\BibitemShut {NoStop}%
\bibitem [{\citenamefont {Kim}\ and\ \citenamefont {Sayama}(2017)}]{Kim2017}%
  \BibitemOpen
  \bibfield  {author} {\bibinfo {author} {\bibnamefont {Kim}, \bibfnamefont
  {M.}}\ and\ \bibinfo {author} {\bibnamefont {Sayama}, \bibfnamefont {H.}},\
  }\bibfield  {title} {\enquote {\bibinfo {title} {Predicting stock market
  movements using network science: an information theoretic approach},}\ }\href
  {\doibase 10.1007/s41109-017-0055-y} {\bibfield  {journal} {\bibinfo
  {journal} {Applied Network Science}\ }\textbf {\bibinfo {volume} {2}},\
  \bibinfo {pages} {35} (\bibinfo {year} {2017})}\BibitemShut {NoStop}%
\bibitem [{\citenamefont {Laib}\ \emph {et~al.}(2018)\citenamefont {Laib},
  \citenamefont {Golay}, \citenamefont {Telesca},\ and\ \citenamefont
  {Kanevski}}]{Laib2018c}%
  \BibitemOpen
  \bibfield  {author} {\bibinfo {author} {\bibnamefont {Laib}, \bibfnamefont
  {M.}}, \bibinfo {author} {\bibnamefont {Golay}, \bibfnamefont {J.}}, \bibinfo
  {author} {\bibnamefont {Telesca}, \bibfnamefont {L.}}, \ and\ \bibinfo
  {author} {\bibnamefont {Kanevski}, \bibfnamefont {M.}},\ }\bibfield  {title}
  {\enquote {\bibinfo {title} {Multifractal analysis of the time series of
  daily means of wind speed in complex regions},}\ }\href {\doibase
  https://doi.org/10.1016/j.chaos.2018.02.024} {\bibfield  {journal} {\bibinfo
  {journal} {Chaos, Solitons \& Fractals}\ }\textbf {\bibinfo {volume} {109}},\
  \bibinfo {pages} {118 -- 127} (\bibinfo {year} {2018})}\BibitemShut {NoStop}%
\bibitem [{\citenamefont {Laib}, \citenamefont {Telesca},\ and\ \citenamefont
  {Kanevski}(2018{\natexlab{a}})}]{Laib2018b}%
  \BibitemOpen
  \bibfield  {author} {\bibinfo {author} {\bibnamefont {Laib}, \bibfnamefont
  {M.}}, \bibinfo {author} {\bibnamefont {Telesca}, \bibfnamefont {L.}}, \ and\
  \bibinfo {author} {\bibnamefont {Kanevski}, \bibfnamefont {M.}},\ }\bibfield
  {title} {\enquote {\bibinfo {title} {Long-range fluctuations and
  multifractality in connectivity density time series of a wind speed
  monitoring network},}\ }\href {\doibase 10.1063/1.5022737} {\bibfield
  {journal} {\bibinfo  {journal} {Chaos: An Interdisciplinary Journal of
  Nonlinear Science}\ }\textbf {\bibinfo {volume} {28}},\ \bibinfo {pages}
  {033108} (\bibinfo {year} {2018}{\natexlab{a}})}\BibitemShut {NoStop}%
\bibitem [{\citenamefont {Laib}, \citenamefont {Telesca},\ and\ \citenamefont
  {Kanevski}(2018{\natexlab{b}})}]{Laib2018a}%
  \BibitemOpen
  \bibfield  {author} {\bibinfo {author} {\bibnamefont {Laib}, \bibfnamefont
  {M.}}, \bibinfo {author} {\bibnamefont {Telesca}, \bibfnamefont {L.}}, \ and\
  \bibinfo {author} {\bibnamefont {Kanevski}, \bibfnamefont {M.}},\ }\bibfield
  {title} {\enquote {\bibinfo {title} {Periodic fluctuations in
  correlation-based connectivity density time series: Application to wind
  speed-monitoring network in switzerland},}\ }\href {\doibase
  https://doi.org/10.1016/j.physa.2017.11.081} {\bibfield  {journal} {\bibinfo
  {journal} {Physica A: Statistical Mechanics and its Applications}\ }\textbf
  {\bibinfo {volume} {492}},\ \bibinfo {pages} {1555 -- 1569} (\bibinfo {year}
  {2018}{\natexlab{b}})}\BibitemShut {NoStop}%
\bibitem [{\citenamefont {Lancichinetti}, \citenamefont {Fortunato},\ and\
  \citenamefont {Radicchi}(2008)}]{Lancichinetti2008}%
  \BibitemOpen
  \bibfield  {author} {\bibinfo {author} {\bibnamefont {Lancichinetti},
  \bibfnamefont {A.}}, \bibinfo {author} {\bibnamefont {Fortunato},
  \bibfnamefont {S.}}, \ and\ \bibinfo {author} {\bibnamefont {Radicchi},
  \bibfnamefont {F.}},\ }\bibfield  {title} {\enquote {\bibinfo {title}
  {Benchmark graphs for testing community detection algorithms},}\ }\href
  {\doibase 10.1103/PhysRevE.78.046110} {\bibfield  {journal} {\bibinfo
  {journal} {Phys. Rev. E}\ }\textbf {\bibinfo {volume} {78}},\ \bibinfo
  {pages} {046110} (\bibinfo {year} {2008})}\BibitemShut {NoStop}%
\bibitem [{\citenamefont {Newman}(2003)}]{Newman2003}%
  \BibitemOpen
  \bibfield  {author} {\bibinfo {author} {\bibnamefont {Newman}, \bibfnamefont
  {M.}},\ }\bibfield  {title} {\enquote {\bibinfo {title} {The structure and
  function of complex networks},}\ }\href {\doibase 10.1137/S003614450342480}
  {\bibfield  {journal} {\bibinfo  {journal} {SIAM Review}\ }\textbf {\bibinfo
  {volume} {45}},\ \bibinfo {pages} {167--256} (\bibinfo {year}
  {2003})}\BibitemShut {NoStop}%
\bibitem [{\citenamefont {Newman}(2004)}]{Modu}%
  \BibitemOpen
  \bibfield  {author} {\bibinfo {author} {\bibnamefont {Newman}, \bibfnamefont
  {M.~E.~J.}},\ }\bibfield  {title} {\enquote {\bibinfo {title} {Analysis of
  weighted networks},}\ }\href {\doibase 10.1103/PhysRevE.70.056131} {\bibfield
   {journal} {\bibinfo  {journal} {Phys. Rev. E}\ }\textbf {\bibinfo {volume}
  {70}},\ \bibinfo {pages} {056131} (\bibinfo {year} {2004})}\BibitemShut
  {NoStop}%
\bibitem [{\citenamefont {de~Oliveira~Santos}, \citenamefont {Stosic},\ and\
  \citenamefont {Stosic}(2012)}]{DEOLIVEIRASANTOS2012}%
  \BibitemOpen
  \bibfield  {author} {\bibinfo {author} {\bibnamefont {de~Oliveira~Santos},
  \bibfnamefont {M.}}, \bibinfo {author} {\bibnamefont {Stosic}, \bibfnamefont
  {T.}}, \ and\ \bibinfo {author} {\bibnamefont {Stosic}, \bibfnamefont
  {B.~D.}},\ }\bibfield  {title} {\enquote {\bibinfo {title} {Long-term
  correlations in hourly wind speed records in {P}ernambuco, {B}razil},}\
  }\href@noop {} {\bibfield  {journal} {\bibinfo  {journal} {Physica A:
  Statistical Mechanics and its Applications}\ }\textbf {\bibinfo {volume}
  {391}},\ \bibinfo {pages} {1546 -- 1552} (\bibinfo {year}
  {2012})}\BibitemShut {NoStop}%
\bibitem [{\citenamefont {Pierini}, \citenamefont {Lovallo},\ and\
  \citenamefont {Telesca}(2012)}]{PIERINI2012}%
  \BibitemOpen
  \bibfield  {author} {\bibinfo {author} {\bibnamefont {Pierini}, \bibfnamefont
  {J.~O.}}, \bibinfo {author} {\bibnamefont {Lovallo}, \bibfnamefont {M.}}, \
  and\ \bibinfo {author} {\bibnamefont {Telesca}, \bibfnamefont {L.}},\
  }\bibfield  {title} {\enquote {\bibinfo {title} {Visibility graph analysis of
  wind speed records measured in central {A}rgentina},}\ }\href@noop {}
  {\bibfield  {journal} {\bibinfo  {journal} {Physica A: Statistical Mechanics
  and its Applications}\ }\textbf {\bibinfo {volume} {391}},\ \bibinfo {pages}
  {5041 -- 5048} (\bibinfo {year} {2012})}\BibitemShut {NoStop}%
\bibitem [{\citenamefont {Pons}\ and\ \citenamefont {Latapy}(2005)}]{walkt}%
  \BibitemOpen
  \bibfield  {author} {\bibinfo {author} {\bibnamefont {Pons}, \bibfnamefont
  {P.}}\ and\ \bibinfo {author} {\bibnamefont {Latapy}, \bibfnamefont {M.}},\
  }\bibfield  {title} {\enquote {\bibinfo {title} {Computing communities in
  large networks using random walks},}\ }in\ \href@noop {} {\emph {\bibinfo
  {booktitle} {Computer and Information Sciences - ISCIS 2005}}},\ \bibinfo
  {editor} {edited by\ \bibinfo {editor} {\bibfnamefont {p.}~\bibnamefont
  {Yolum}}, \bibinfo {editor} {\bibfnamefont {T.}~\bibnamefont
  {G{\"u}ng{\"o}r}}, \bibinfo {editor} {\bibfnamefont {F.}~\bibnamefont
  {G{\"u}rgen}}, \ and\ \bibinfo {editor} {\bibfnamefont {C.}~\bibnamefont
  {{\"O}zturan}}}\ (\bibinfo  {publisher} {Springer Berlin Heidelberg},\
  \bibinfo {address} {Berlin, Heidelberg},\ \bibinfo {year} {2005})\ pp.\
  \bibinfo {pages} {284--293}\BibitemShut {NoStop}%
\bibitem [{\citenamefont {{R Core Team}}(2018)}]{lanR}%
  \BibitemOpen
  \bibfield  {author} {\bibinfo {author} {\bibnamefont {{R Core Team}},},\
  }\href@noop {} {\emph {\bibinfo {title} {R: A Language and Environment for
  Statistical Computing}}},\ \bibinfo {organization} {R Foundation for
  Statistical Computing},\ \bibinfo {address} {Vienna, Austria} (\bibinfo
  {year} {2018})\BibitemShut {NoStop}%
\bibitem [{\citenamefont {Reichardt}\ and\ \citenamefont
  {Bornholdt}(2006)}]{SPG}%
  \BibitemOpen
  \bibfield  {author} {\bibinfo {author} {\bibnamefont {Reichardt},
  \bibfnamefont {J.}}\ and\ \bibinfo {author} {\bibnamefont {Bornholdt},
  \bibfnamefont {S.}},\ }\bibfield  {title} {\enquote {\bibinfo {title}
  {Statistical mechanics of community detection},}\ }\href {\doibase
  10.1103/PhysRevE.74.016110} {\bibfield  {journal} {\bibinfo  {journal} {Phys.
  Rev. E}\ }\textbf {\bibinfo {volume} {74}},\ \bibinfo {pages} {016110}
  (\bibinfo {year} {2006})}\BibitemShut {NoStop}%
\bibitem [{\citenamefont {Rosvall}\ and\ \citenamefont
  {Bergstrom}(2007)}]{infom}%
  \BibitemOpen
  \bibfield  {author} {\bibinfo {author} {\bibnamefont {Rosvall}, \bibfnamefont
  {M.}}\ and\ \bibinfo {author} {\bibnamefont {Bergstrom}, \bibfnamefont
  {C.~T.}},\ }\bibfield  {title} {\enquote {\bibinfo {title} {An
  information-theoretic framework for resolving community structure in complex
  networks},}\ }\href {\doibase 10.1073/pnas.0611034104} {\bibfield  {journal}
  {\bibinfo  {journal} {Proceedings of the National Academy of Sciences}\
  }\textbf {\bibinfo {volume} {104}},\ \bibinfo {pages} {7327--7331} (\bibinfo
  {year} {2007})}\BibitemShut {NoStop}%
\bibitem [{\citenamefont {Rousseeuw}(1987)}]{Silhouettes1987}%
  \BibitemOpen
  \bibfield  {author} {\bibinfo {author} {\bibnamefont {Rousseeuw},
  \bibfnamefont {P.~J.}},\ }\bibfield  {title} {\enquote {\bibinfo {title}
  {Silhouettes: A graphical aid to the interpretation and validation of cluster
  analysis},}\ }\href {\doibase https://doi.org/10.1016/0377-0427(87)90125-7}
  {\bibfield  {journal} {\bibinfo  {journal} {Journal of Computational and
  Applied Mathematics}\ }\textbf {\bibinfo {volume} {20}},\ \bibinfo {pages}
  {53 -- 65} (\bibinfo {year} {1987})}\BibitemShut {NoStop}%
\bibitem [{\citenamefont {Steinhaeuser}, \citenamefont {Chawla},\ and\
  \citenamefont {Ganguly}(2009)}]{Steinhaeuser2009}%
  \BibitemOpen
  \bibfield  {author} {\bibinfo {author} {\bibnamefont {Steinhaeuser},
  \bibfnamefont {K.}}, \bibinfo {author} {\bibnamefont {Chawla}, \bibfnamefont
  {N.~V.}}, \ and\ \bibinfo {author} {\bibnamefont {Ganguly}, \bibfnamefont
  {A.~R.}},\ }\bibfield  {title} {\enquote {\bibinfo {title} {An exploration of
  climate data using complex networks},}\ }in\ \href@noop {} {\emph {\bibinfo
  {booktitle} {Proceedings of the Third International Workshop on Knowledge
  Discovery from Sensor Data}}}\ (\bibinfo  {publisher} {ACM},\ \bibinfo {year}
  {2009})\ pp.\ \bibinfo {pages} {23--31}\BibitemShut {NoStop}%
\bibitem [{\citenamefont {Telesca}, \citenamefont {Lovallo},\ and\
  \citenamefont {Kanevski}(2016)}]{Telesca2016}%
  \BibitemOpen
  \bibfield  {author} {\bibinfo {author} {\bibnamefont {Telesca}, \bibfnamefont
  {L.}}, \bibinfo {author} {\bibnamefont {Lovallo}, \bibfnamefont {M.}}, \ and\
  \bibinfo {author} {\bibnamefont {Kanevski}, \bibfnamefont {M.}},\ }\bibfield
  {title} {\enquote {\bibinfo {title} {Power spectrum and multifractal
  detrended fluctuation analysis of high-frequency wind measurements in
  mountainous regions},}\ }\href@noop {} {\bibfield  {journal} {\bibinfo
  {journal} {Applied Energy}\ }\textbf {\bibinfo {volume} {162}},\ \bibinfo
  {pages} {1052 -- 1061} (\bibinfo {year} {2016})}\BibitemShut {NoStop}%
\bibitem [{\citenamefont {Treiber}, \citenamefont {Heinermann},\ and\
  \citenamefont {Kramer}(2016)}]{Treiber2016}%
  \BibitemOpen
  \bibfield  {author} {\bibinfo {author} {\bibnamefont {Treiber}, \bibfnamefont
  {N.~A.}}, \bibinfo {author} {\bibnamefont {Heinermann}, \bibfnamefont {J.}},
  \ and\ \bibinfo {author} {\bibnamefont {Kramer}, \bibfnamefont {O.}},\
  }\enquote {\bibinfo {title} {Wind power prediction with machine learning},}\
  in\ \href@noop {} {\emph {\bibinfo {booktitle} {Computational
  Sustainability}}}\ (\bibinfo  {publisher} {Springer International
  Publishing},\ \bibinfo {year} {2016})\ Chap.\ \bibinfo {chapter} {Wind Power
  Prediction with Machine Learning}, pp.\ \bibinfo {pages} {13--29}\BibitemShut
  {NoStop}%
\bibitem [{\citenamefont {Tsonis}\ and\ \citenamefont
  {Roebber}(2004)}]{Tsonis2004}%
  \BibitemOpen
  \bibfield  {author} {\bibinfo {author} {\bibnamefont {Tsonis}, \bibfnamefont
  {A.}}\ and\ \bibinfo {author} {\bibnamefont {Roebber}, \bibfnamefont {P.}},\
  }\bibfield  {title} {\enquote {\bibinfo {title} {The architecture of the
  climate network},}\ }\href@noop {} {\bibfield  {journal} {\bibinfo  {journal}
  {Physica A: Statistical Mechanics and its Applications}\ }\textbf {\bibinfo
  {volume} {333}},\ \bibinfo {pages} {497 -- 504} (\bibinfo {year}
  {2004})}\BibitemShut {NoStop}%
\bibitem [{\citenamefont {Tsonis}\ and\ \citenamefont
  {Swanson}(2008)}]{Tsonis2008}%
  \BibitemOpen
  \bibfield  {author} {\bibinfo {author} {\bibnamefont {Tsonis}, \bibfnamefont
  {A.~A.}}\ and\ \bibinfo {author} {\bibnamefont {Swanson}, \bibfnamefont
  {K.~L.}},\ }\bibfield  {title} {\enquote {\bibinfo {title} {Topology and
  predictability of el {N}i\~no and la {N}i\~na networks},}\ }\href@noop {}
  {\bibfield  {journal} {\bibinfo  {journal} {Phys. Rev. Lett.}\ }\textbf
  {\bibinfo {volume} {100}},\ \bibinfo {pages} {228--502} (\bibinfo {year}
  {2008})}\BibitemShut {NoStop}%
\bibitem [{\citenamefont {Tsonis}, \citenamefont {Swanson},\ and\ \citenamefont
  {Roebber}(2006)}]{Tsonis2006}%
  \BibitemOpen
  \bibfield  {author} {\bibinfo {author} {\bibnamefont {Tsonis}, \bibfnamefont
  {A.~A.}}, \bibinfo {author} {\bibnamefont {Swanson}, \bibfnamefont {K.~L.}},
  \ and\ \bibinfo {author} {\bibnamefont {Roebber}, \bibfnamefont {P.~J.}},\
  }\bibfield  {title} {\enquote {\bibinfo {title} {What do networks have to do
  with climate?}}\ }\href@noop {} {\bibfield  {journal} {\bibinfo  {journal}
  {Bulletin of the American Meteorological Society}\ }\textbf {\bibinfo
  {volume} {87}},\ \bibinfo {pages} {585--595} (\bibinfo {year}
  {2006})}\BibitemShut {NoStop}%
\bibitem [{\citenamefont {Yamasaki}, \citenamefont {Gozolchiani},\ and\
  \citenamefont {Havlin}(2008)}]{Yamasaki2008}%
  \BibitemOpen
  \bibfield  {author} {\bibinfo {author} {\bibnamefont {Yamasaki},
  \bibfnamefont {K.}}, \bibinfo {author} {\bibnamefont {Gozolchiani},
  \bibfnamefont {A.}}, \ and\ \bibinfo {author} {\bibnamefont {Havlin},
  \bibfnamefont {S.}},\ }\bibfield  {title} {\enquote {\bibinfo {title}
  {Climate networks around the globe are significantly affected by el
  {N}i\~no},}\ }\href@noop {} {\bibfield  {journal} {\bibinfo  {journal} {Phys.
  Rev. Lett.}\ }\textbf {\bibinfo {volume} {100}},\ \bibinfo {pages} {228501}
  (\bibinfo {year} {2008})}\BibitemShut {NoStop}%
\bibitem [{\citenamefont {Yang}, \citenamefont {Algesheimer},\ and\
  \citenamefont {Tessone}(2016)}]{Yang2016}%
  \BibitemOpen
  \bibfield  {author} {\bibinfo {author} {\bibnamefont {Yang}, \bibfnamefont
  {Z.}}, \bibinfo {author} {\bibnamefont {Algesheimer}, \bibfnamefont {R.}}, \
  and\ \bibinfo {author} {\bibnamefont {Tessone}, \bibfnamefont {C.~J.}},\
  }\bibfield  {title} {\enquote {\bibinfo {title} {A comparative analysis of
  community detection algorithms on artificial networks},}\ }\href {\doibase
  10.1038/srep30750} {\bibfield  {journal} {\bibinfo  {journal} {Scientific
  reports}\ }\textbf {\bibinfo {volume} {6}},\ \bibinfo {pages} {30750}
  (\bibinfo {year} {2016})}\BibitemShut {NoStop}%
\end{thebibliography}%

\cleardoublepage

\end{document}